\documentclass[10pt,conference]{IEEEtran}
\usepackage{amsmath}
\usepackage{algorithm}
\usepackage{url}
\usepackage{graphicx}
\usepackage{multirow}
\usepackage{multicol}
\usepackage{verbatim}
\usepackage{todonotes}
\usepackage{amsfonts}
\usepackage{indentfirst}
\usepackage[noend]{algpseudocode}
\usepackage{caption}
\usepackage{subcaption}
\usepackage{array}
\usepackage{bm}
\usepackage{color}
\usepackage{verbatim}
\usepackage{epstopdf}
\usepackage{xspace}
\usepackage{cite}

\ifCLASSINFOpdf
\else
\fi

\begin{document}
\title{Catching Anomalous Distributed Photovoltaics: An Edge-based Multi-modal Anomaly Detection}

%\author{\IEEEauthorblockN{Michael Shell}
%\IEEEauthorblockA{School of Electrical and\\Computer Engineering\\
%Georgia Institute of Technology\\
%Atlanta, Georgia 30332--0250\\
%Email: http://www.michaelshell.org/contact.html}
%\and
%\IEEEauthorblockN{Homer Simpson}
%\IEEEauthorblockA{Twentieth Century Fox\\
%\and
%\IEEEauthorblockN{James Kirk\\ and Montgomery Scott}
%\IEEEauthorblockA{Starfleet Academy\\
%San Francisco, California 96678--2391\\


\author{\IEEEauthorblockN{Devu Manikantan Shila\IEEEauthorrefmark{2}, Kin Gwn Lore\IEEEauthorrefmark{2},Tianshu Wei\IEEEauthorrefmark{3}, Teems Lovett\IEEEauthorrefmark{2} and Yu Cheng\IEEEauthorrefmark{4}}
\IEEEauthorblockA{\IEEEauthorrefmark{2}United Technologies Research Center, Hartford, Connecticut 06108 \\
 Email:\{manikad, lorek,lovettt\}@utrc.utc.com}
\IEEEauthorblockA{\IEEEauthorrefmark{3}University of California, Riverside, California, 92507, \\
Email: twei002@ucr.edu}
\IEEEauthorblockA{\IEEEauthorrefmark{4}Illinois Institute of Technology, Chicago, Illinois 60616 \\
 Email:cheng@iit.edu}
}


\maketitle

\begin{abstract}
A significant challenge in energy system cyber security is
the current inability to detect cyber-physical attacks targeting and
originating from distributed grid-edge devices such as photovoltaics (PV) panels, smart flexible loads, and electric vehicles. Cyber grid defenders lack
the necessary algorithms and other detection capabilities to
distinguish between normal operations, cyber-attacks, and other
exceptional circumstances. We address this concern by designing and developing a distributed, multi-modal anomaly detection approach that can sense the health of the device and the electric power grid from the edge. This is realized by exploiting unsupervised machine learning algorithms on multiple sources of time-series data such as voltage magnitude and phase angles from the area where the device interconnects with the power grid, power injected to and absorbed from the grid, and power quality of the edge device, fusing these multiple local observations and flagging anomalies when a deviation from the normal behavior is observed.
\vspace{5pt}\\
We particularly focus on the cyber-physical threats to the distributed PVs that has the potential to cause local disturbances or grid instabilities by creating supply-demand mismatch, reverse power flow conditions etc. We use an open source power system simulation tool called GridLAB-D, loaded with  real smart home and solar datasets to simulate the smart grid scenarios and to illustrate the impact of PV attacks on the power system. Various attacks targeting PV panels that create voltage fluctuations, reverse power flow etc were designed and performed. We observe that while individual unsupervised learning algorithms such as OCSVMs, Corrupt RF and PCA surpasses in identifying particular attack type, PCA with Convex Hull outperforms all algorithms in identifying all designed attacks with a true positive rate of 83.64\% and an accuracy of 95.78\%. Our key insight is that due to the heterogeneous nature of the distribution grid and the uncertainty in the type of the attack being launched, relying on single mode of information for defense can lead to increased false alarms and missed detection rates as one can design attacks to hide within those uncertainties and remain stealthy. 
\end{abstract}

\IEEEpeerreviewmaketitle
\section{Introduction}
\label{sec:intro}
Power grid utilities are currently moving towards
a smarter grid using new technologies such as smart meters,
real-time pricing, demand side flexibility and distributed energy
resources (DERs - flexible loads, distributed generation, energy storage, and electric vehicles). These advanced smart grid technologies will be 
designed and deployed to improve the operations and the efficacy
of the electrical grid; nevertheless there is a cost associated with these technologies:
{\em increased system exposure and expanded attack surface}~\cite{Kushner:2013, DER:2017}. These
exposures provide new cyber security vulnerabilities that have
the potential to be exploited by attackers. Examples include an attacker who potentially takes control of a large number of smart meters and command them to simultaneously perform a disconnect and reconnect, causing grid frequency variations that may be significant enough to generate a disruption of service~\cite{Smartgridawareness:2015} or, an attacker who alters data at the smart meter leading to billing and trending inaccuracies~\cite{Costache:CND2011}. 

Grid operators and balancing authorities typically do not have sufficient visibility into the secondary distribution and end-use levels at individual residential and commercial buildings or DER installations. Such limited visibility may be utilized by an attacker to launch attacks on the grid through the end-user devices such as compromised smart meters, building automation systems or renewable energy sources~\cite{Rasche:IGCTD2015,Smartgridawareness:2015,Nutaro:ORNL2014}. Moreover, the existence of uncertainties
in the distribution grid provide ample opportunities for an
adversary to execute attacks without being
detected. A large scale compromise of a significant number of building automation technology, such as networked energy management systems may potentially cause a disruption in energy service by exploiting the remote shut-off functionality and causing an imbalance in the grid~\cite{howtohack:2016}. The impact of such an attack could be exacerbated if the attack took place during peak hours of electricity delivery and at a large scale. Cyber defenders currently lack the visibility to detect and quickly respond to such an attack.  

To detect targeted cyber-attacks and achieve attack resiliency, there is a requirement for continuous monitoring of  DERs and their interactions with the electrical grid in real-time. This work specifically focuses on the cyber security of distributed photovoltaics (PVs), where the objective of an adversary is to exploit the stochastic power generation/consumption characteristics and vulnerabilities in control system environment to create supply-demand mismatch and reverse power flow conditions that can lead to local disturbances or grid instabilities~\cite{Solar}. A  handful of efforts have been proposed for various analyses in power systems using data-driven techniques \cite{Gardner:PES2006, Nutaro:ORNL2014, Srikantha:ISGT2015, Liu:DSC2016, Amini:ISGT2015, Mousavian:GW2015} such as distributed state estimation, topology error identification, smart meter anomaly detection, malicious data detection, faulty PV detection, intrusion detection of automated generation control and SCADA systems,  etc.  However, there are significant gaps in the techniques and technology to detect potentially destabilizing cyber-physical attacks emerging from the DERs (or grid edge devices) and report those observations in a timely manner for enhanced situational awareness. The one that is close to ours is \cite{jamie}, where authors employ $\mu$PMUs for situational awareness in the distribution grid; essentially, anomaly detection of distributed automation equipments by leveraging CUSUM techniques on the physics based features derived from $\mu$PMUs. An anomaly is flagged by monitoring the change in the voltage-current correlations which in turn leads to increased false positives since such changes in correlations also occurs with normal grid disturbances (e.g., voltage sag). 

As a solution to this problem, to the best of our knowledge, we first design and develop a {\em distributed, edge-based multi-modal anomaly detection scheme} that continuously observes PV and its interactions with the electric power grid by exploiting unsupervised learning algorithms on multiple sources of time-series data: {\em voltage magnitude and phase angles at the feeder node, power injected to and absorbed from the grid, power characteristics of PVs} and then fuse these multiple observations and flag anomalies when a deviation from the normal behavior is observed. 

We use various unsupervised regression and classification techniques to design a robust anomaly detection framework. The use of unsupervised techniques enables us to ``detect any anomalous deviations'' without relying on the knowledge of specific attack patterns. On the other hand, insufficient knowledge of attack pattens will also make the model prone to false alarms~\cite{Shila:HS2015}. We amortize the cost of false alarms and missed detection by utilizing multiple sources of information that provide cues on the behavior of the device as well as the its interactions with the grid. To model the smart grid scenarios and further illustrate the impact of anomalous PVs on the power systems, we use an open source power system simulation tool called GridLAB-D~\cite{chassin2008gridlab}, loaded with realistic smart home and solar datasets from UMass Trace Repository~\cite{barker2012smart, chen2016sunspot}. Various attacks such as {\em reverse power flow}, {\em disconnect}, {\em power curtailment} and {\em volt-var} at different attack penetration levels (25\% to 100\%) that exploit the normal behavior of distributed generation to cause voltage fluctuations, supply-demand mismatch, grid instabilities were designed and implemented in GridLAB-D.  We make the following observations from our simulations and analyses:
\begin{itemize}
\item Each machine learning algorithm surpasses in identifying a particular type of attack. For e.g., one class SVMs \cite{chen2001one} and PCA with Convex Hull~\cite{jolliffe2002principal} detect {\em volt-var} and {\em power curtailment} attacks with (accuracy, F1) scores of (99.81\%, 98.83\%) and (95.00\%, 70.75\%), respectively. Corrupt Random Forest~\cite{liaw2002classification} detects {\em reverse power flow} and {\em disconnect} attacks with (accuracy, F1) scores of (95\%, 76.07\%) and (97.22\%, 85.59\%) respectively. 

\item PCA with Convex Hull proved to be the most robust algorithm to detect all four attacks with a precision, recall, F1, accuracy and receiver operating curve area of 70.02\%, 83.64\%, 76.22\%, 95.78\% and 90.24\%, respectively. 

\item Heterogeneous nature of the distribution grid power generation and demand limits one to use single modality for anomaly detection that may lead to increased false alarms and missed detection. For e.g., {\em reverse power flow} attacks that causes increased flow of energy back to the substation cannot be identified from PV power characteristics but can be detected by monitoring for anomalous changes in the power exported back to the grid. The uncertainties in the power generation/consumption also explain the rationale behind the low F1 score in identifying particular attack types such as {\em power curtailment attacks} that can hide within those fluctuating load profiles and remain stealthy. Since these attacks behave like ``normal events'', no noticeable impact are imposed on the power grid characteristics. 

\item Anomalous variations of feeder measurements such as voltage magnitudes and phase angles provide a valuable source to identify large scale attacks from the edge. Using these measurements, we were able to identify attacks with a true positive rate of 30\% to 40\% for various attack penetration levels. The low detection rates stem from the use of low fidelity data obtained from the GridLAB-D that in turn fails to capture voltage-current transients and due to the heterogeneous nature of the distribution grid itself.
\end{itemize}
The rest of the paper is organized as follows: In Section~\ref{sec:approach}, we present the system and adversary model. Section~\ref{sec:detection} presents the edge-based multi-modal anomaly detection framework, machine learning algorithms, and dataset generation from GridLAB-D simulations.  In section~\ref{sec:results}, we present the performance of the anomaly detection framework. Finally, section~\ref{sec:conclusion} concludes this work. 
\section{System and Adversary Model}
\label{sec:approach}
\subsection{Distributed Photovoltaics model and Issues}
\begin{figure}[h]
    \centering
    \includegraphics[width=\columnwidth]{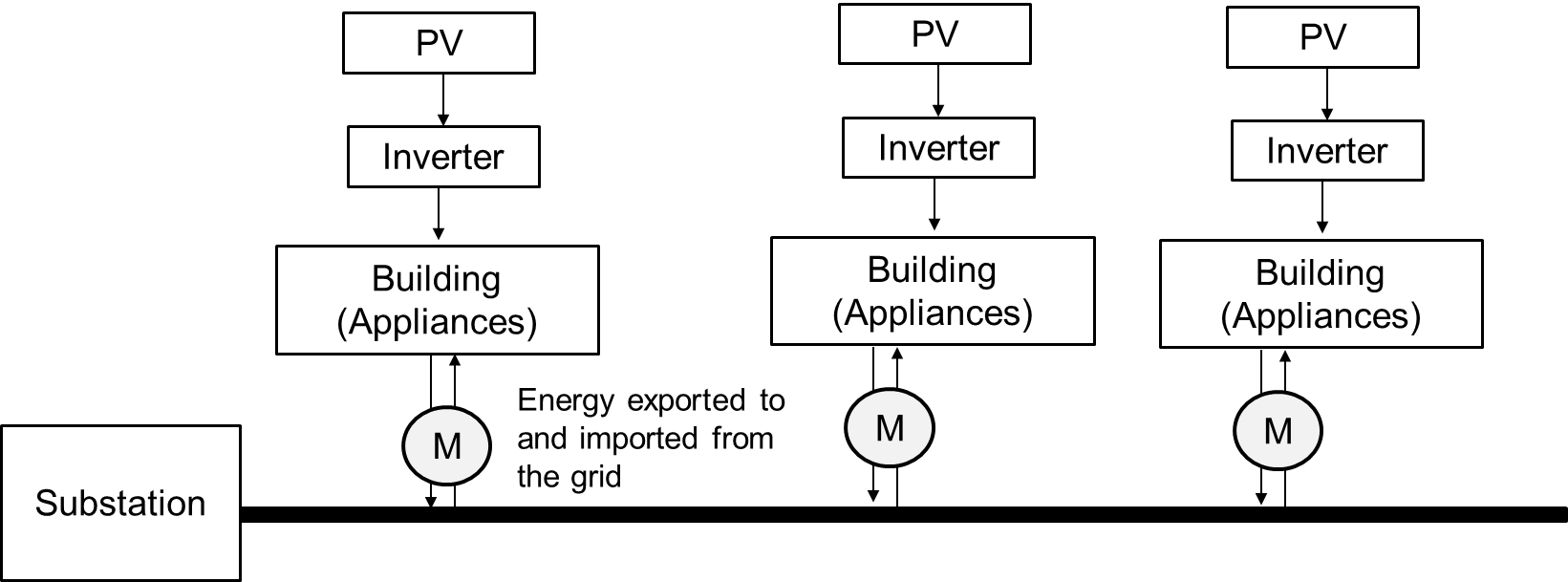}
    \caption{System model consisting of distributed photovoltaics and their integration with the grid}
    \label{figure:system}
\end{figure}

We consider a group of buildings (residential or commercial) with rooftop embedded PVs connected to the distribution feeder network (medium/low voltage network) via a bi-directional meter (see Figure~\ref{figure:system}). Each building consists of a
Energy Management System (EMS) that monitors, manages and controls the energy produced by the PV and the energy consumed by the appliances within the buildings. The energy consumed as well as the excess energy exported is monitored with the help of a bi-directional meter (\mbox{M}). The output power of PV, which is dependent on the incident solar radiation, is converted to AC power to use within a building by a PV inverter. Excess energy produced by the PV is exported back to the electric power grid, which typically occurs when the supply of PV
generated  power is greater than the power
demand of the building loads (typically referred to as reverse power flow).  
This situation arise around midday in dwellings
when the PV generation is high, as the available solar
energy is highest in the middle of day, and the
dwellings electricity consumption is low as the
occupants may be, for example, out at work. In general, distributed generation brings two advantages:
(i) reduces the need to buy energy from the power grid, thereby reducing the feeder load demand and the electricity bills for the customer by utilizing the PV generated power directly by the buildings loads during the peak time period;  (ii) enables the grid operators to meet the feeder load demands by using the unwanted PV power exported to the grid. 

In spite of these advantages, increased penetration of PVs on the distribution grid present several opportunities
and challenges for power distribution utilities~\cite{Solar}. Major
adverse impacts of high PV penetration are on system voltages. For e.g., steady state voltage rises with reverse power flow or
voltage fluctuations caused by rapid changes in the PV output.
Existing efforts propose to offset the voltage rise in distribution networks
by exploiting the inherent reactive power capability of the PV
inverters that can inject or absorb reactive power as needed~\cite{Hess:SGT2016} .
Moreover, unlike traditional generation plants, PVs do not have rotating parts, and their output is highly sensitive to solar radiation which causes undesirable voltage fluctuations that can affect the operation of protection equipment. 
In order to rectify the fluctuations in the PV output caused either due
to a large variance of solar irradiance (voltage rises) or due to the cloud cover (voltage drops), ramp rate control
strategies are proposed~\cite{Alam:EC2014}. 
\begin{figure*}[htb]
    \centering
    \includegraphics[width=0.7\textwidth,trim={96 32 96 0}]{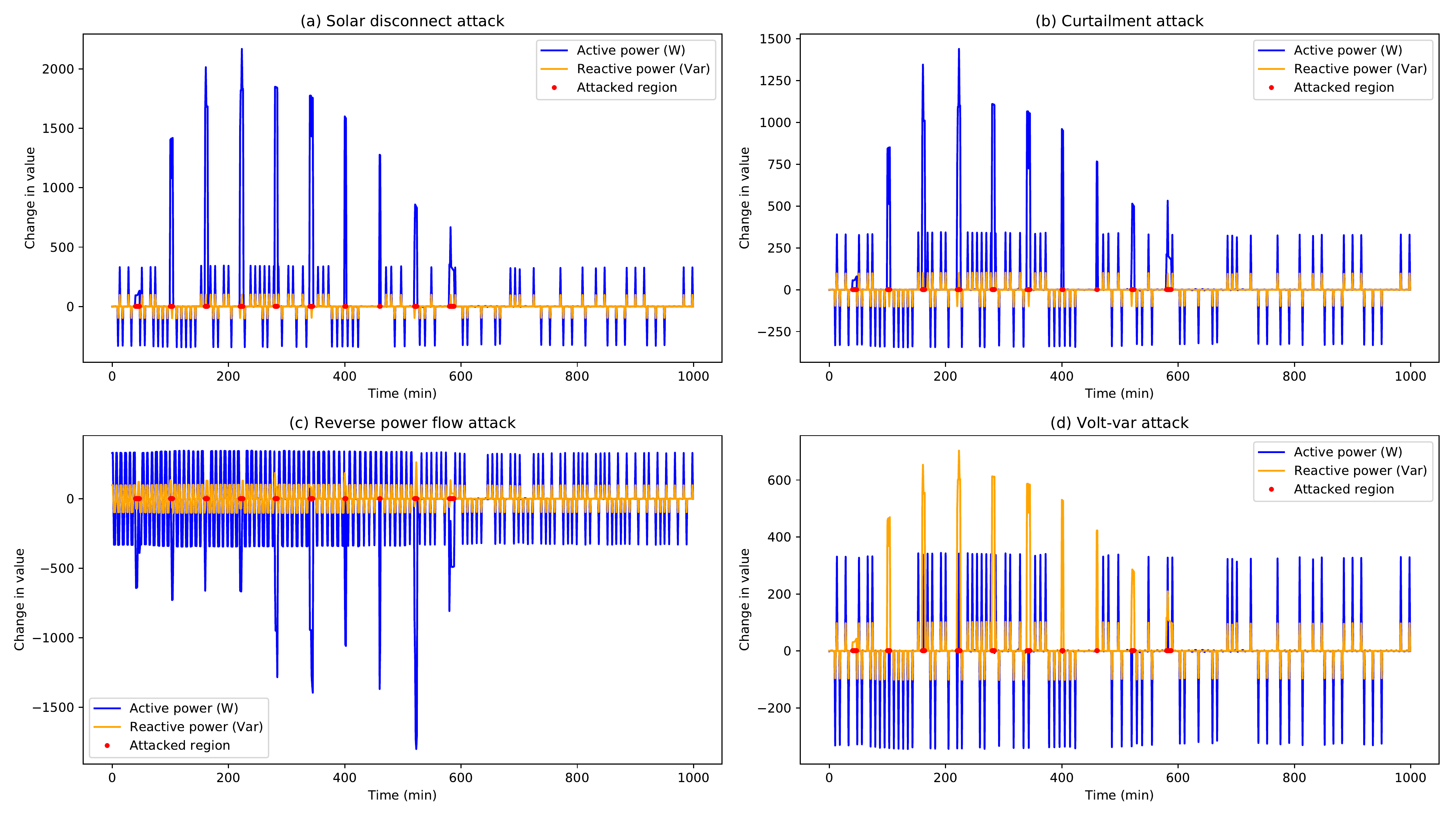}
    \caption{Active power and reactive power from GridLAB-D simulations for four attacks. The red points denote the point at which the attacks are launched and the change in value indicate the impact of an attack. Attacks are performed at regular intervals in a continuous fashion. The y-axis is depicted using different scales to project the attack impact}
    \label{figure_attack}
\end{figure*}
\subsection{Adversary model}
\label{sec:adversary}
We assume an adversary whose
goal is to cause local outages, equipment damages or grid
instabilities by controlling the PVs connected to a distribution
network via a \textbf{cyber attack} (exploiting vulnerable
EMS, local or external communication networks, web services, authorization and
authentication schemes, communication protocols etc.) or a \textbf{physical attack} (e.g., access the system and change the control strategies or the parameters). 
The impact of an attack depends on the number of PVs
compromised, the location of the PVs (e.g., close to the
substation or away from the substation), the condition, configuration
and characteristics of the distribution feeder and the
type of the attack. We design four types of attacks on PV systems and is discussed below: 
\begin{itemize}
\item \textit{Disconnect attack:} In this attack, an adversary gains control to a large number of PVs and issues a mass disconnect command which disconnects the PV from the house during normal operational hours. This attack may be launched at regular intervals in a continuous fashion (e.g., disconnect-connect-disconnect) or in a random manner, say disconnect under heavy load conditions. Such an attack will force the building to draw more energy from the grid, specifically under heavy load conditions. This will incur increased power demand from the power grid and consequently, overloading of the feeders leading to frequency/voltage violations and system instabilities. Disconnection of PVs will also affect the grid planning and operations as it prevents customer from selling energy back to the grid when needed. 

\item \textit{Power curtailment attack:} The objective of an adversary is to manipulate the maximum active power output of the PV inverter. By tampering the control algorithms or the control parameters (e.g., change the power factor or ramp rate via EMS), attacker tunes the maximum power output of the PV inverter to a certain percentage of the current available solar power (e.g., setting to half of the real power). Although these attacks may not bring significant impact like the solar disconnect attacks, they are hard to detect as these attacks can hide within those fluctuating PV output power. 

\item \textit{Volt-var attack:} As said earlier,  PV inverters will be exploited in future to inject certain amount of reactive power to help maintain a stable voltage in distribution networks. We assume that the attacker has exploited the weaknesses in the control algorithms or the communication protocols to change the control parameters related to the  inverter. Adversary particularly manipulates the power factor of the inverter to arbitrarily inject different level of reactive power, which in turn can greatly affect the voltage magnitude and phase angles in the grid.

\item \textit{Reverse power flow attack:} Many residential houses have been equipped with Internet-connected smart appliances which enable demand response techniques (e.g., control the load of the devices by utilities or via EMS by the customer). The objective of an attacker is to exploit the weaknesses in EMS itself or the Internet connect smart appliances to turn them off or the circuit breakers to reduce the house load demand. This will significantly increase the reverse power flow injected back to the grid under heavy power generation conditions (e.g., when the solar power generation is high during the daytime) and the feeder will in turn start exporting power to the neighboring feeders. By controlling a number of houses installed with the PVs, these attacks can negatively affect protection coordination and operation of line voltage regulators, leading to voltage fluctuations and instabilities. 
\end{itemize}
We conduct experiments in GridLAB-D to demonstrate the impact of these attacks on the electric power grid. Figure \ref{figure_attack} illustrates the active and reactive power obtained from the bi-directional meter under the four attack modes. All attacks were conducted during heavy power generation periods (i.e., daytime) to highlight the negative effects.  

Following observations were made from Figure \ref{figure_attack}: (a) Attacks such as {\em volt-var} can be identified by monitoring anomalous deviations in reactive power; on the other hand, attacks such as {\em disconnect} and {\em power curtailment} mainly causes variations to the active power; (b) {\em Disconnect} attacks can be easily identified as active power drops to zero. On the other hand, one of the practical challenge in detecting this attack is to distinguish between {\em disconnect} attacks and normal events where a user disconnects and reconnects the PV; (c) Due to the fast variation of the solar output power and also the fact that the {\em power curtailment attacks} manipulates the PV power factor, it is hard to detect these attacks as they can hide within those output variations; (d) {\em Reverse power flow} attacks are hard to identify by merely observing the output power of PV as they are typically conducted by disconnecting the house loads. The negative ``change in value'' in the figure indicates the amount of power exported back to the grid during the attack, which is obtained from the bi-directional meter. 

\begin{figure}[htbp]
\centering
% disconnect attack
\begin{subfigure}{0.23\textwidth}
  \centering
  \includegraphics[width=4.7cm]{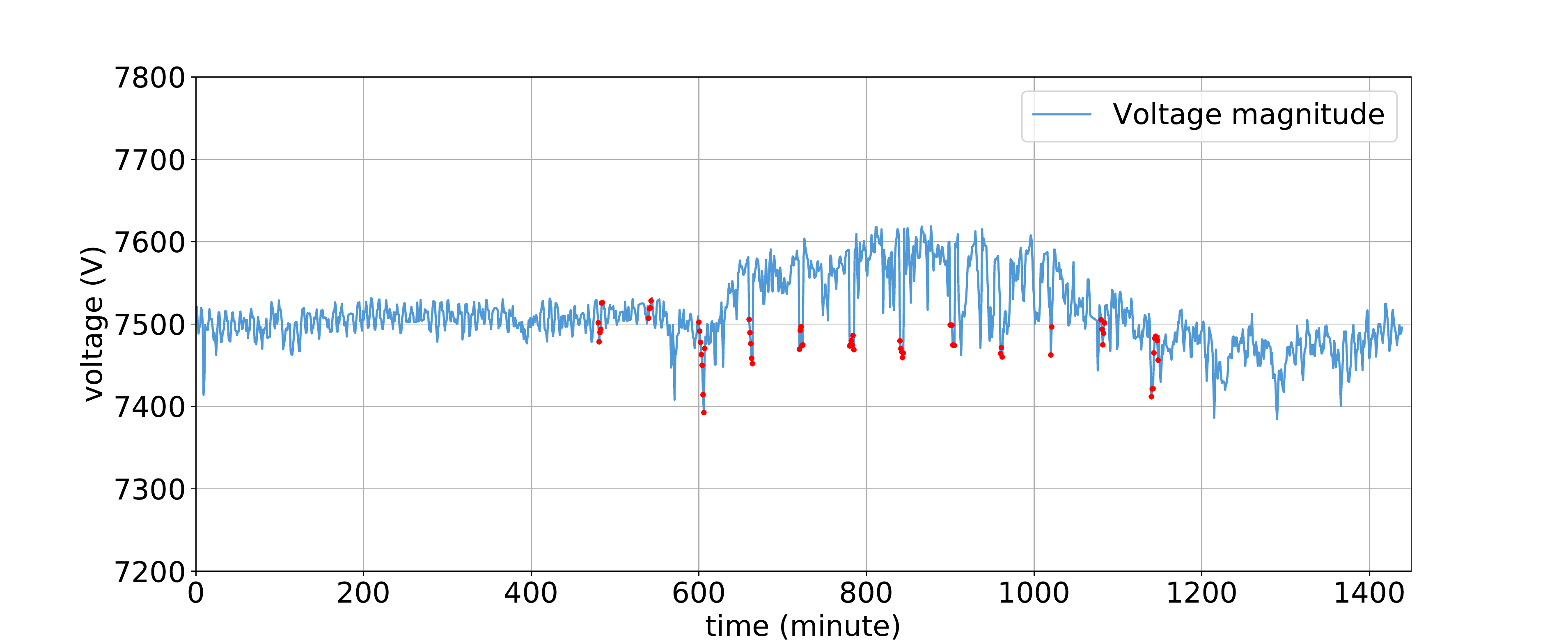}
  \caption{Voltage maginitude during disconnect attack}
  \label{figure_voltage_disconnect}
\end{subfigure} \hspace{0in}
\begin{subfigure}{0.23\textwidth}
  \centering
  \includegraphics[width=4.7cm]{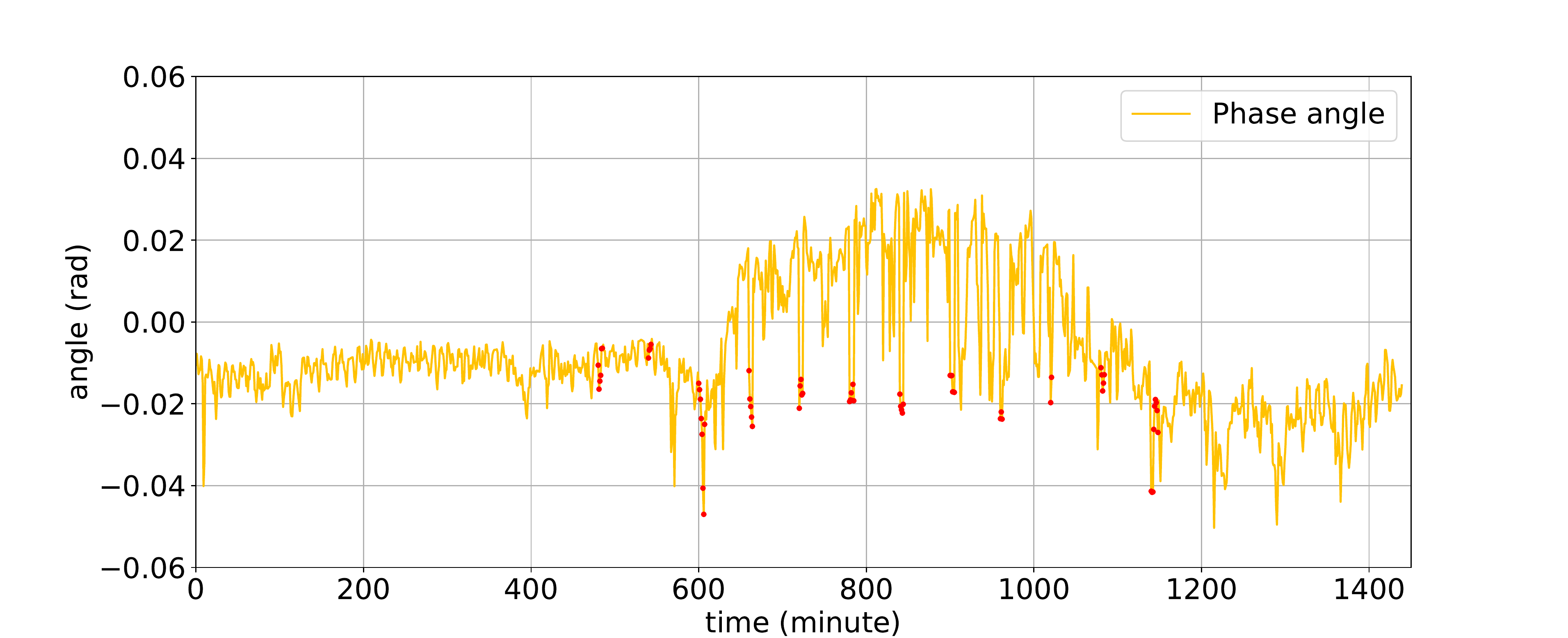}
  \caption{Phase angle during disconnect attack}
  \label{figure_angle_disconnect}
\end{subfigure}
\\ % curtailment attack
\begin{subfigure}{0.23\textwidth}
  \centering
  \includegraphics[width=4.7cm]{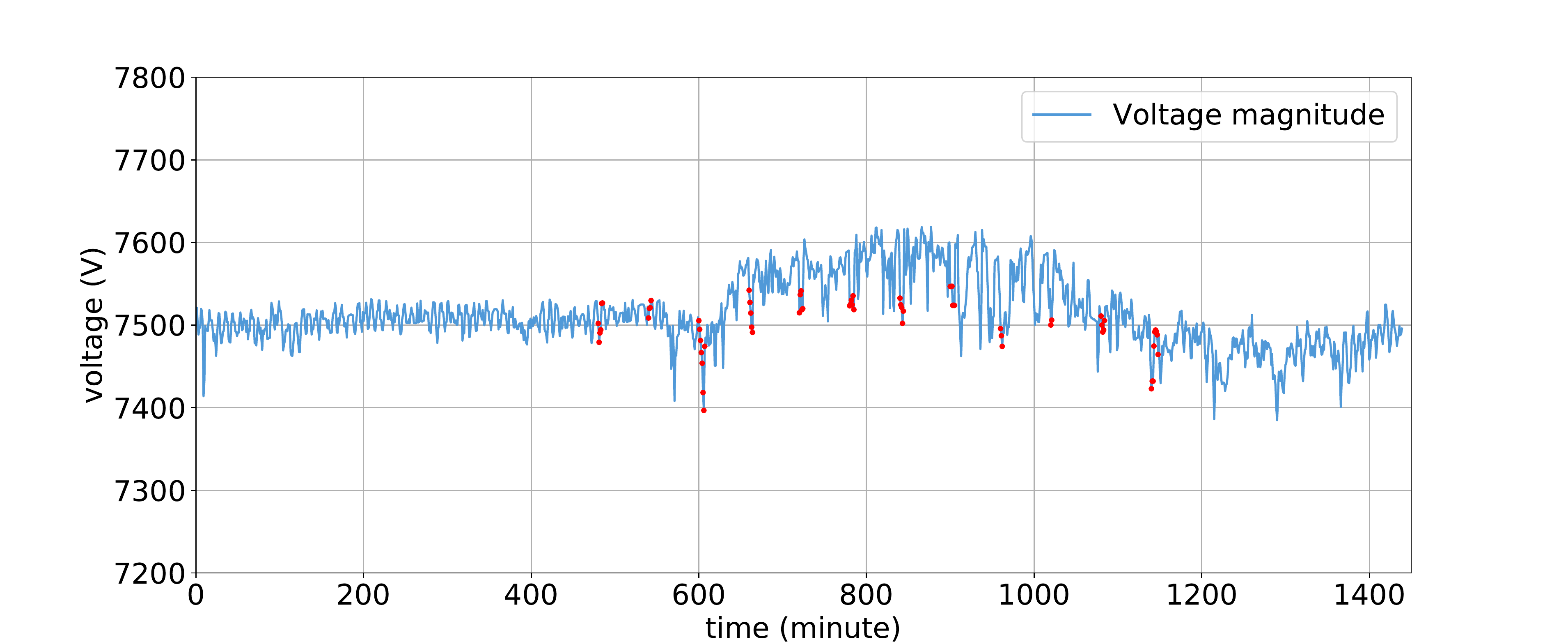}
  \caption{Voltage maginitude during curtailment attack}
  \label{figure_voltage_curtailment}
\end{subfigure} \hspace{0in}
\begin{subfigure}{0.23\textwidth}
  \centering
  \includegraphics[width=4.7cm]{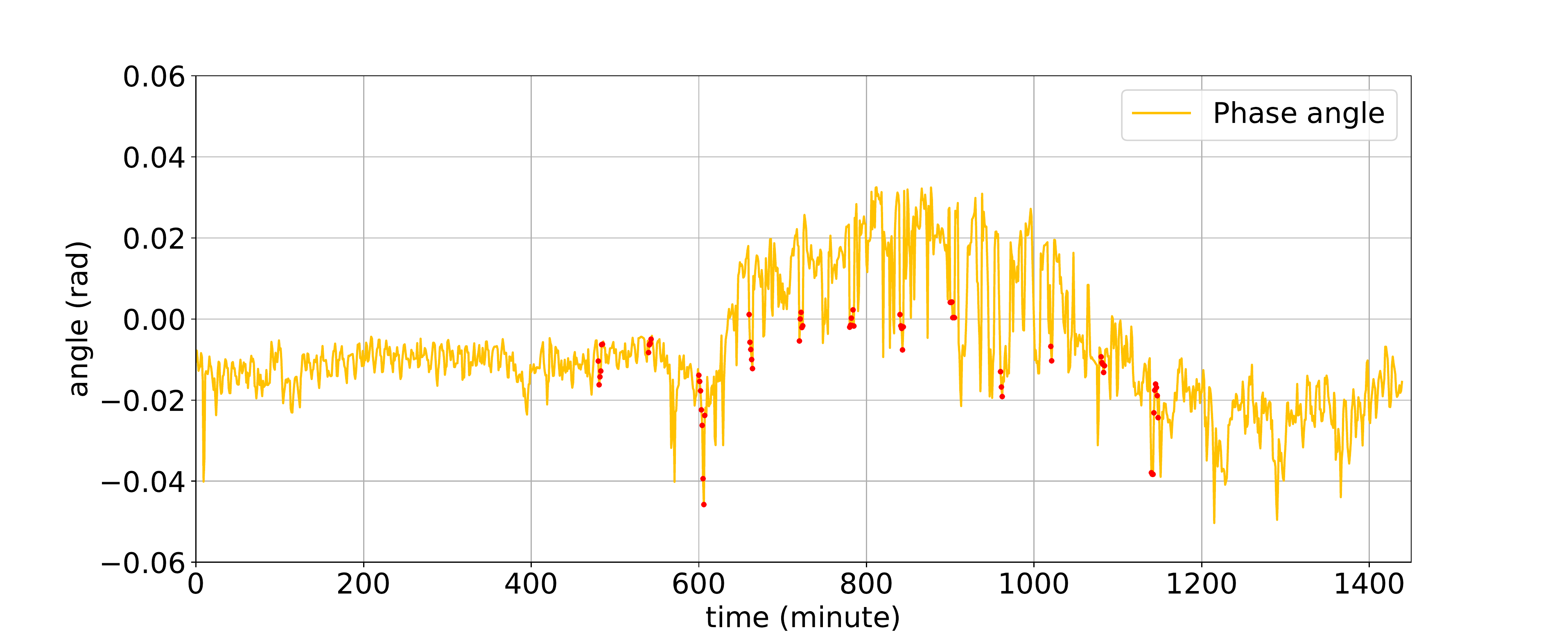}
  \caption{Phase angle during curtailment attack}
  \label{figure_angle_curtailment}
\end{subfigure}
\\ % var attack
\begin{subfigure}{0.23\textwidth}
  \centering
  \includegraphics[width=4.7cm]{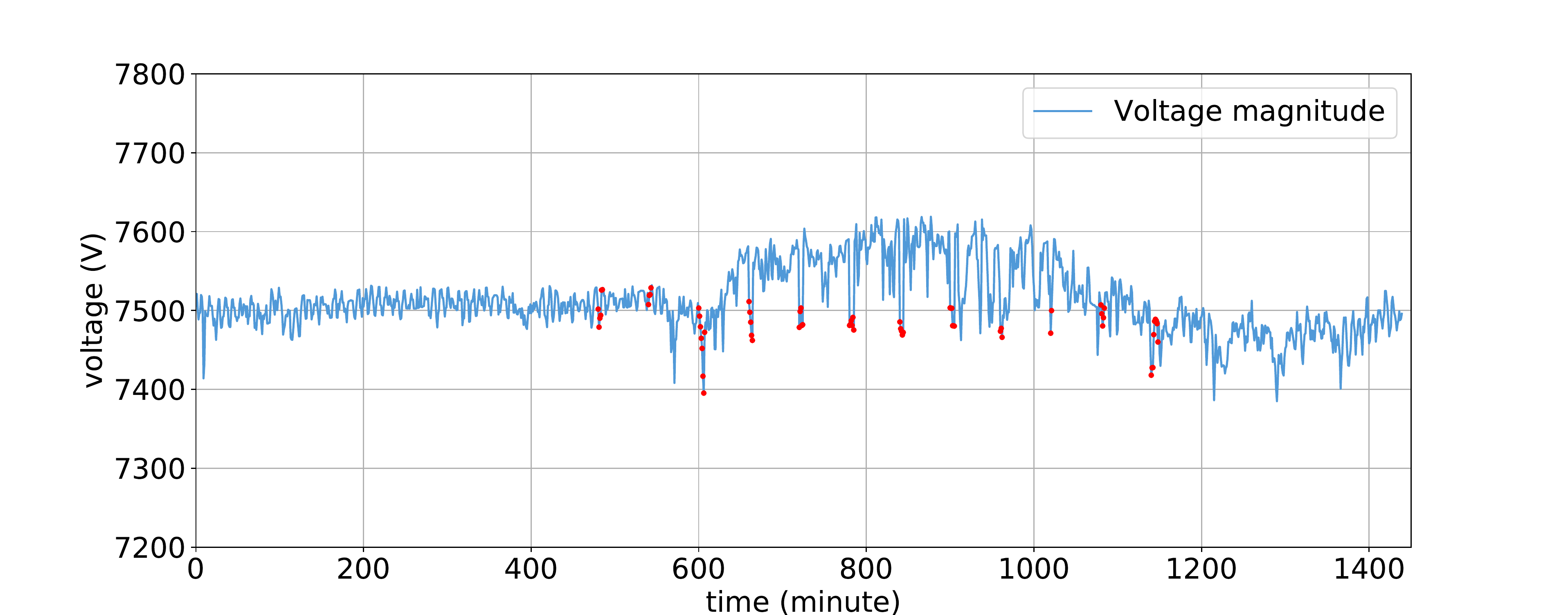}
  \caption{Voltage maginitude during volt-var attack}
  \label{figure_voltage_var}
\end{subfigure} \hspace{0in}
\begin{subfigure}{0.23\textwidth}
  \centering
  \includegraphics[width=4.7cm]{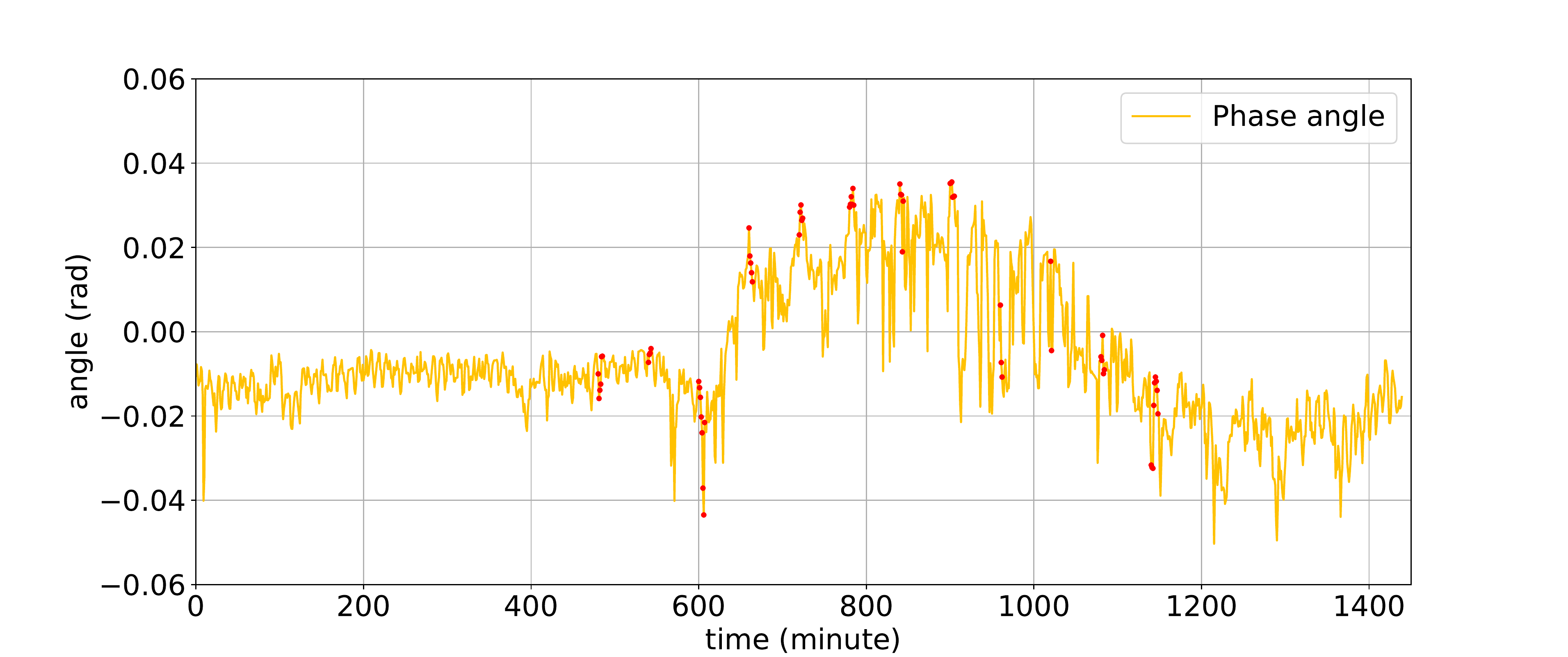}
  \caption{Phase angle during volt-var attack}
  \label{figure_angle_var}
\end{subfigure}
\\ % load attack
\begin{subfigure}{0.23\textwidth}
  \centering
  \includegraphics[width=4.7cm]{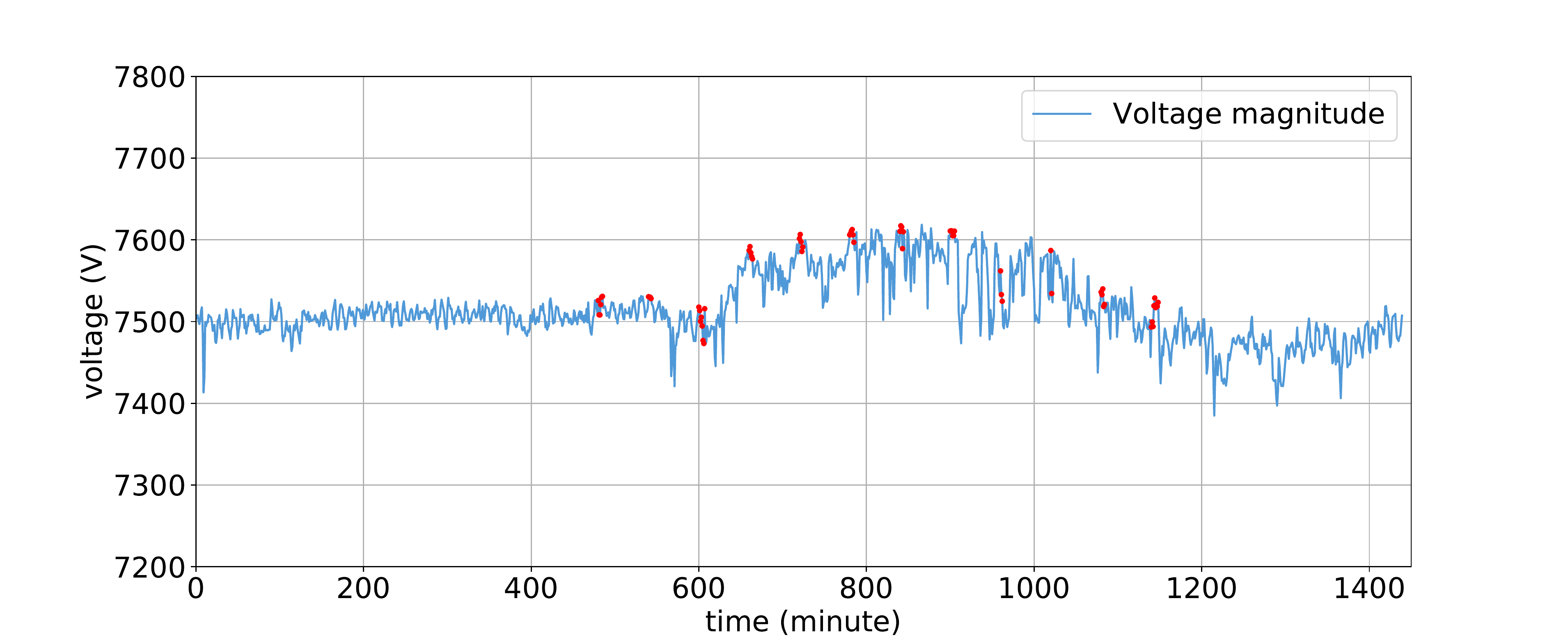}
  \caption{Voltage maginitude during reverse power attack}
  \label{figure_voltage_load}
\end{subfigure} \hspace{0in}
\begin{subfigure}{0.23\textwidth}
  \centering
  \includegraphics[width=4.7cm]{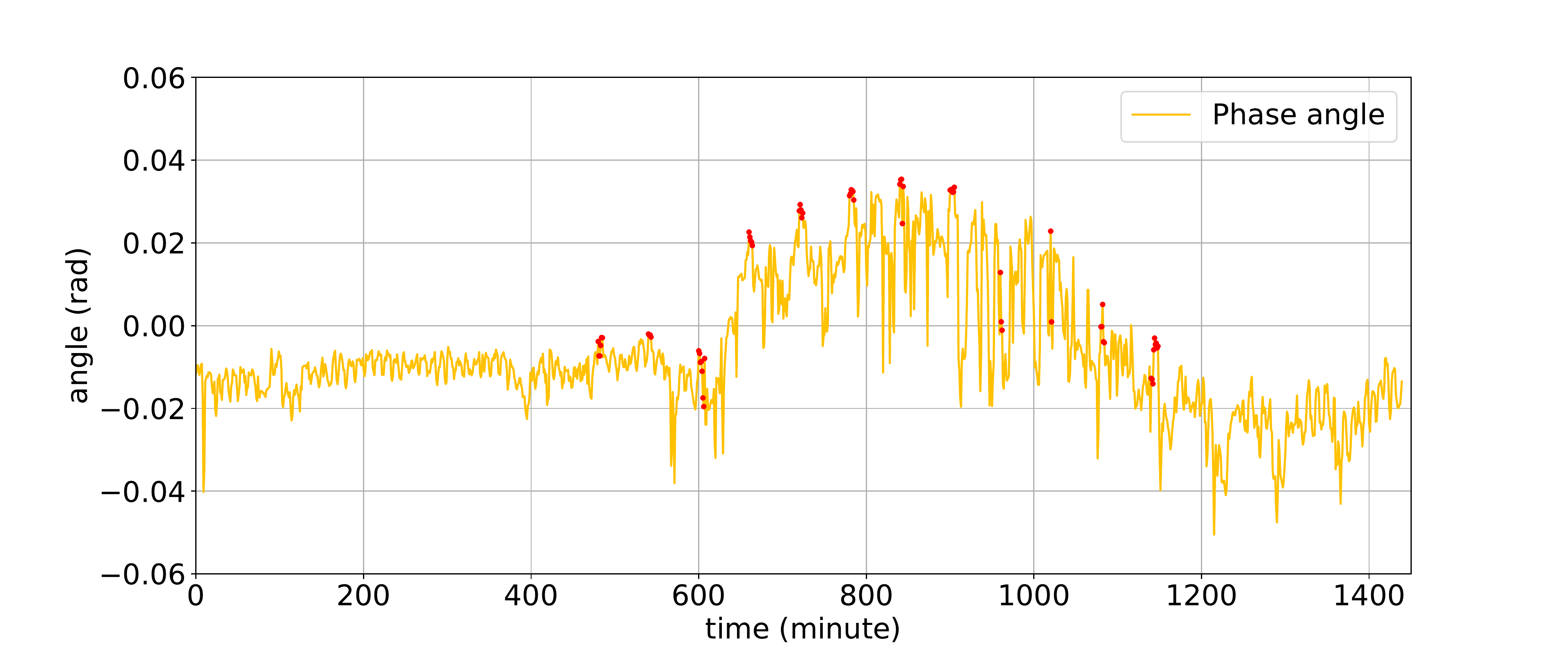}
  \caption{Phase angles during reverse power flow attack }
  \label{figure_angle_load}
\end{subfigure}
\caption{Voltage magnitude and phase angle from simulated PMUs at the point of interconnect under various types of attacks. The red points denote the magnitude and phase angles when an attack occurs.}
\label{figure_attack_pmu}
\end{figure}

As mentioned earlier, when a large number of systems are compromised (e.g., attacker performing power curtailment attacks on 75\% of the connected systems), it will create noticeable traces on the voltage in the distribution feeder at the point where the building/device interfaces with the grid. One of the attacker strategy is to control a number of devices and manipulate the loads in a manner so that it will be undetected by the anomaly detection algorithms at the edge (say, when the change in value is within the threshold). Although these attacks may not be observable through device features, the feeder voltage magnitude and phase angles will be affected (especially for weak feeders) which in turn can be observed and measured using additional sensors such as $\mu$-PMU (micro Phasor Measurement Units e.g., PQube3 from PSL~\cite{PQube3:2016}). Figure \ref{figure_attack_pmu} shows the voltage magnitude and phase angle when the four attacks are performed. 

One could observe different behavior of the feeder voltage magnitude and phase angles with regard to the four attack modes. (a) In the event of {\em disconnect} attack and {\em curtailment} attack, both voltage magnitude and phase angle  drops from nominal values as more power is drawn from the grid to meet the load demands; (b) During the {\em volt-var} attack, injection of more reactive power into the grid leads to drop in voltage magnitude and rise in phase angle. Typically, voltage increases with the reactive power. In our case, the reactive power is increasing the voltage at the edge which results in a voltage drop at the feeder node due to the voltage balance; (c) In the {\em reverse power flow} attack, both the magnitude and phase angle of feeder voltage rises as more real power is injected into the grid. By observing those changes in the voltage magnitude and phase angle, one could identify the type of the attacks conduced; however, in our analysis, we observed that due to the uncertainties in the distribution grid (e.g., stochastic nature of the loads), voltage fluctuations are normal. This in turn makes it hard to assertively claim if the fluctuation observed was due to a normal event or an attack. This also implies that missed detection of anomalies which lie within the normal behavior is not a concern if the grid voltages/frequencies are stable and are within the limits. 
\section{Edge-based multi-modal anomaly detection framework}
\label{sec:detection}
In this section, we present the multi-modal anomaly detection framework, techniques to learn the anomaly detection models, and simulations performed to learn and evaluate the performance of our framework. 
\subsection{Smart grid simulation and dataset generation}
\begin{figure}[htb]
    \centering
    \includegraphics[width=1\columnwidth]{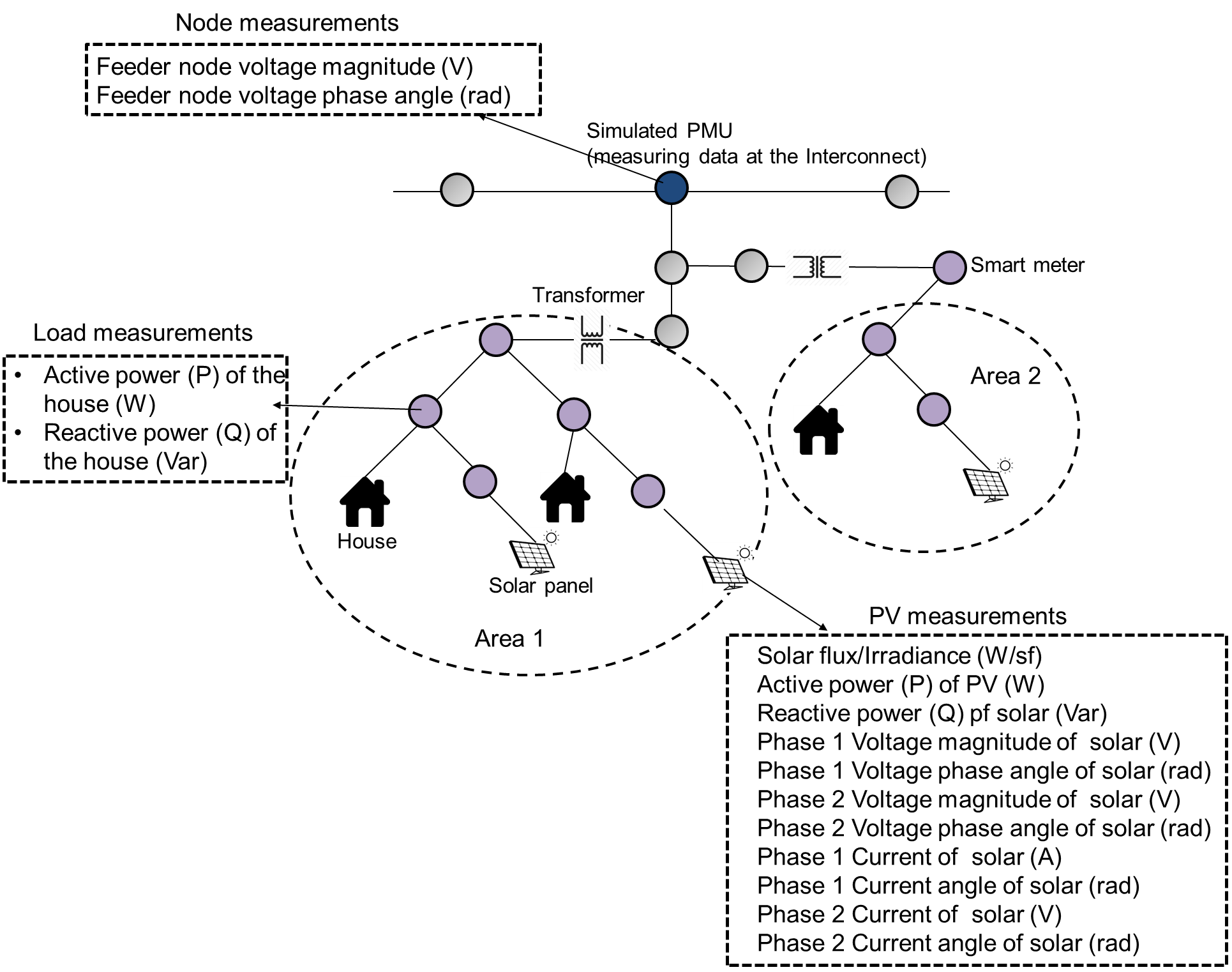}
    \caption{The system model in Figure 1 simulated in GridLAB-D. Radial distribution feeder models with house, PV panels and sensors (PMU, smart meters) and the dataset generated from simulations (node, PV and load measurements) are also illustrated}
    \label{figure:feeder}
\end{figure}

We run the simulations using GridLAB-D, which is an open-source software
platform developed by the U.S. Department of Energy
at the Pacific Northwest National Laboratory (PNNL) for the
simulation of electric power distribution systems. 
The GridLAB-D prototypical feeder model were used to represent the radial distribution network. They represent the fundamental characteristics of radial
distribution feeders found in the U.S., based on 575 distribution
feeders from 151 separate substations from different utilities
across the U.S. Each prototypical feeder is characterized as
belonging to one of five U.S. climate regions, by primary distribution
voltage level, and other features. The taxonomy feeder models are provided as part of the GridLAB-D
software package.  The feeders have been
modeled with high fidelity from the substation down to the
individual customer meters, including detailed end-use load
representations (heating, ventilation, and air-conditioning, and
various other constant impedance, current, and power loads). In this work, we use the moderate suburban and rural feeder network (R1-1247-1) with 12.5 kV and 7152 kVA ratings for simulations. As these original prototypical feeder models only represent the topology of the distribution network, we developed python-based scripts to automatically add residential house and rooftop PVs to the original feeder (.glm) models to represent the load demand and distributed power generation, respectively, as shown in Figure \ref{figure:feeder}. 
\begin{itemize}
\item {\em Residential house loads}: To capture realistic load demands from residential homes, we used the Smart dataset from UMass Trace Repository~\cite{barker2012smart}. We especially used the high resolution data from three homes, described as home A, B and C in\cite{barker2012smart}. Houses are randomly assigned to the transformer based on its power rating and are loaded with home A, B or C load profiles in a random manner. All the houses are connected to the grid via a smart meter, which measures various information such as the power consumed or exported by each individual house, voltage-current characteristics, active/reactive power etc. In our simulations, 6000 houses were automatically added to the  R1-1247-1 feeder model. 
\item {\em Rooftop PV}: To generate the PV output, the incident solar radiation based on the solar irradiance sensing data from UMassTraceRepository~\cite{chen2016sunspot} is used. Using a realistic irradiance data enables us to capture the variations in PV output power which is otherwise lacking in data generated using GridLAB-D provided classic solar irradiance formulas. All the PV panels are virtually installed on residential
rooftops according to a PV penetration level parameter,
which is defined as
\begin{align}
& \mbox{solar~penetration} = \frac{\mbox{maximum~solar~power}}{\mbox{total~grid~peak~load}}\label{equ:penetration}
\end{align}
We used a 25\% PV penetration level which is in accordance with the maximum allowed penetration level by the utility. In our simulations, the total peak solar power generation is $8$MW and the average solar power is around $2$-$3$ MW throughout the day. The generation of solar power can be also controlled based on the peak power demand of the house. Additionally, as the location of PVs produce different impacts on the power grid, we developed python-based scripts to automatically deploy the PVs in three separate regions: (a) close to the substation; (b) in the middle of the feeder and (c) away from the substation. From our simulations, we noticed that when the PV units are aways from the substation, grid voltages are more impacted as opposed to the case where PVs are deployed close to the substation. This is due to the fact that radial distribution systems use voltage regulation equipments located at or close to the substations to regulate the voltage. Any voltage fluctuations close to the substation will be stabilized by these regulators. The results presented in this work, hence, are based on the scenarios where PVs are placed away from the substation. 
\end{itemize}

\textbf{Normal data generation.} We run the simulations in GridLAB-D for summer season from April through August to collect the normal data related to power generation, house demand, and voltage magnitude and phase angles. Figure \ref{figure:feeder} shows the data types generated by our simulations. This includes: (a) {\em house load measurements} - the active and reactive power with respect to each house; the active power also captures the power imported from or exported to the grid via the smart meter; (b) {\em PV measurements} - the active power, reactive power, voltage and current profiles from two phases and (c) {\em Node measurements} -  the voltage magnitude and phase angles in the feeder, measured using simulated PMUs.  The PMU data collected from the point of interconnect (the interface between the residential area with buildings and PVs and the grid) inherently captures the behavior of the edge devices. This is especially true when a large number of houses or PV panels are attacked.  

\textbf{Attack data generation.} We use the dataset generated in the month of September to evaluate the performance of anomaly detection algorithms. The house load, PV and node measurements with regard to the four attack modes were generated. 

\subsection{Anomaly detection framework design}
We develop an {\em edge-based distributed detection framework}, where each building is equipped with a multi-modal anomaly detection scheme that locally observes and analyzes both device and grid behavioral signals to  predict the presence of an attack. Behavioral signals identifies an {\em anomaly} by measuring the abnormality of device's current behavior (i.e., current observed data) with its past recurrent behavior. This phase is called the anomaly prediction phase. The past behavior of the device is learned by applying data-driven techniques on various sources of information gathered from the device and the grid interconnect (see the data used to train the algorithms in Figure \ref{figure:feeder}) and is called the training phase. We use the normal dataset generated from April to August to learn the normal behavioral signals. In particular, we learn three anomaly detection models ($m_1, m_2, m_3$) and is provided below. 
\begin{itemize}
\item PV model ($m_1$): This model learns the normal behavior of each solar panel based on the solar flux/irradiance, active and reactive power generation and the voltage-current characteristics and can be used to detect all attacks except {\em reverse power flow} attack. 
\item Load model ($m_2$): This model detects when anomalous amount of power is injected or drawn from the grid (e.g., {\em disconnect, power curtailment, reverse power flow} attacks) by learning the normal pattern of the load  injected into or absorbed from the grid, which in turn is obtained via smart meters. 
\item Node model ($m_3$): This model detects anomalous voltage magnitude and phase angles when a number of devices in the neighborhood is attacked. Such a detection scheme serve as an alternative to sense the health of the grid from the edge, when all the devices and house loads are under the control of an attacker.
\end{itemize}

\textit{Fusion of $m_1, m_2$ and $m_3$}: Each model operates on three different sets of data and produces anomaly scores during inference. Let $\tilde{m}_{e,i,1}, \tilde{m}_{e,i,2}$ and $\tilde{m}_{e,i,3}$ denote the series of normalized anomaly scores of the evaluation set, computed using features from PV measurements, house load measurements, and node measurements, respectively. Our objective is to fuse these scores to produce a combined  score that determines the degree to which the device is ``under attack''. We explored two approaches: the first scheme is to fuse the normalized anomaly measures from these outputs via \textit{linear combination}. Formally, the fused measure becomes:
\[ \tilde{m}_{e,i} = w_1\tilde{m}_{e,i,1} + w_2\tilde{m}_{e,i,2} + w_3\tilde{m}_{e,i,3} \]
where $w_1, w_2$ and $w_3$ are the weight coefficients. The second scheme, referred to as \textit{MostAnomalous}, is to fuse the normalized anomaly measures by relying on the most anomalous measure, i.e:
\[ m_{e,i} = f(\tilde{m}_{e,i,1},\tilde{m}_{e,i,2},\tilde{m}_{e,i,3}) \]
where $f(\cdot)=\text{min}(\cdot)$ for frameworks where lower measures imply possible attacks and $f(\cdot)=\text{max}(\cdot)$ for frameworks where higher measures imply possible attacks. From our simulation results, we observe that the \textit{MostAnomalous} approach works better compared to linear combination of scores. This is due to the fact that each models specializes in different feature sets and may perform well to detect a certain type of attacks (e.g. {\em disconnect} attacks) but not in the other (e.g. {\em volt-var} attacks or {\em reverse power flow}). Hence, using linear combinations may present issues with generalization.

\subsection{Machine learning algorithms}
We leverage various machine learning algorithms to learn the individual models ($m_1, m_2$ and $m_3$) and is described below. For ease of exposition, PVs, house loads and nodes are collectively referred to as ``device', in the sequel'. 
\subsubsection{ Neural networks} 
\begin{figure}[htb]
    \centering
    \includegraphics[width=1.0\columnwidth]{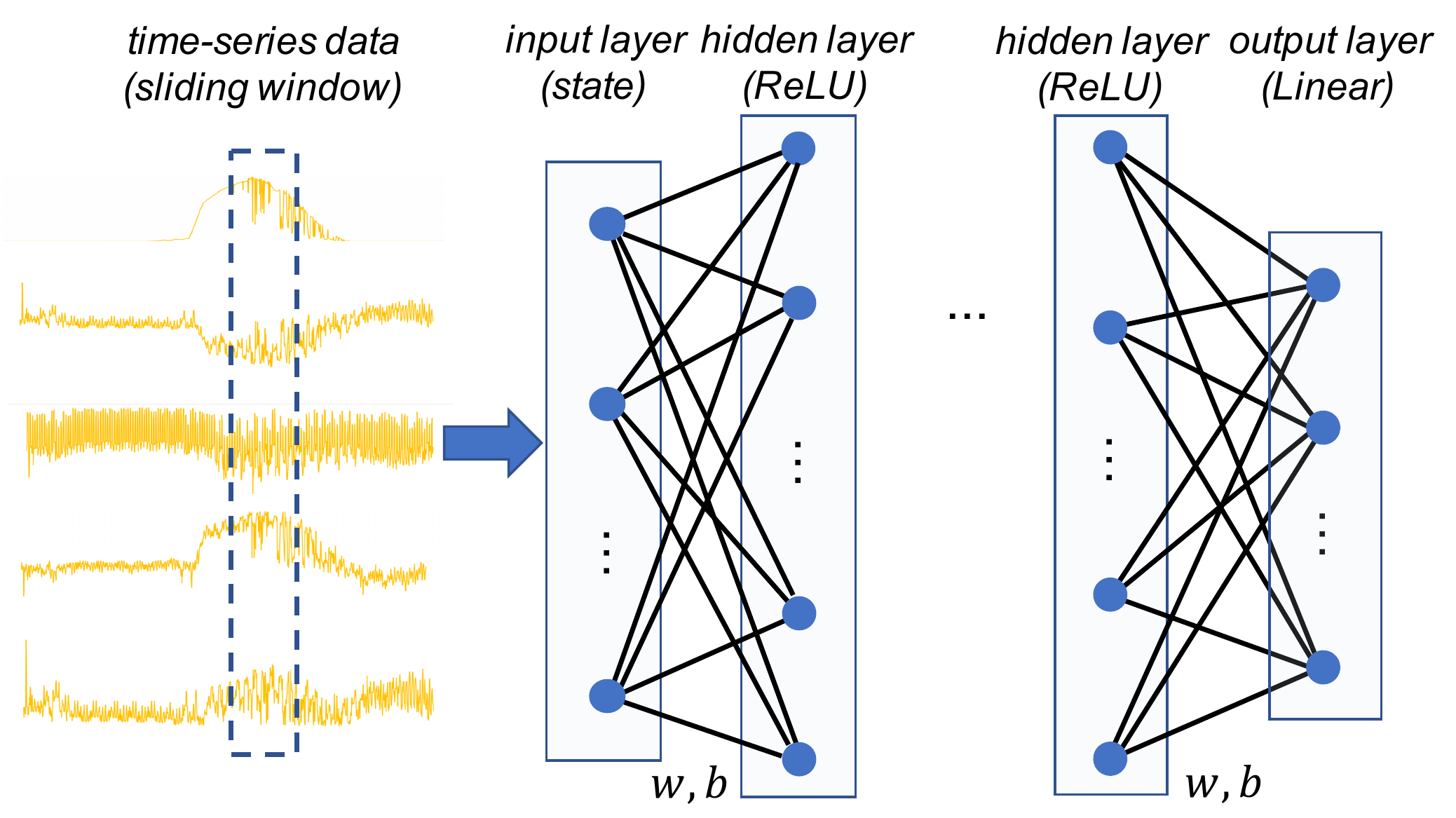}
    \caption{Structure of neural network}
    \label{figure:neural_network}
\end{figure}

We use the artificial neural network to estimate the state of the device at current time stamp based on the recent data measurements. As shown in Figure \ref{figure:neural_network}, there are three hidden layers in the neural network, we use the ReLU (Rectified Linear Unit) as the activation function for hidden layers, while in the output layer, linear activation function is used to estimate the system state at current time stamp. The number of neurons in each hidden layers is $[500,300,100]$. The input of the neural network is the normalized time-series data from different sensors (e.g., solar irradiance, current, voltages and active/reactive power for PV model). As shown in Figure \ref{figure:neural_network}, a sliding window is used to transform the time-series data into the input vector for training the models. During the prediction phase, the estimated system state at current time step is predicted which is compared with the measured system state to calculate the estimation error.  We use multivariate normal distribution to represent the estimation error in the normal scenario (see equation \eqref{equ:multivariate}) where $x$ and $k$ represents the system state estimation error vector and its dimension respectively. The mean $\mu$ and covariance matrix $\Sigma$ is determined from the estimation error in the normal scenario. During the training phase, we derive threshold $\rho$ from the probability density function of the normal distribution. An input time-series data is classified as an anomaly if the computed probability density function is lower than the threshold $\rho$ as shown in Equation \eqref{equ:threshold}.
\begin{align}
& pdf(x) = (2\pi)^{-\frac{1}{2}k}|\Sigma|^{-\frac{1}{2}}e^{-\frac{1}{2}(x-\mu)\Sigma^{-1}(x-\mu)} \label{equ:multivariate}
\end{align}
\begin{align}
& \text{result} = \left\{\begin{array}{ll}
\text{anormaly} & ,pdf(x)<\rho\\
\text{normal} & ,pdf(x)\geq\rho\\
\end{array}\right. \label{equ:threshold}
\end{align}

\subsubsection{Denoising Auto Encoders} An Auto Encoder (AE) is simply a network which tries to reconstruct the input. The key is that AEs first encode the input to a compact vector representation (in the code layer) and then decode to reconstruct the input. The simplest way of forcing the net to discover a compact representation of the data is to have a code layer with fewer dimensions than the input, as in principal component analysis. A Denoising Auto Encoder (DAE) is an auto encoder which attempts to reconstruct a clean target from a noisy input. In this work, we train a stacked DAE to reconstruct the time-series data at each time step using the normal dataset. Similar to neural neural network training presented in (1), we use a Gaussian distribution to model the reconstruction error at the output of auto encoder during the training phase. At the prediction phase, if the probability density function of the average reconstruction error is below a certain threshold, we flag the observation as anomalous.

In our simulations, we use a sliding window length of $15$ minutes to train DAEs and Neural Networks; as our experiments show that $15$-min sliding window achieves the best performance based on our evaluation metrics among different length of sliding windows ($1$-min to $20$-min).

\textbf{Feature engineering.} In algorithms (3)-(7) described below, we normalized the input features to have zero mean and unit variance. While most of the algorithms approach the anomaly detection problem as a classification task, we find no improvement in performance when employing sliding window (e.g. 15 minute windows) or rolling features (rolling mean, range, standard deviations). In most observations, rolling features in such cases either {\em delay the flagging of the anomaly} or {\em unable to flag as normal once the attack has been lifted}, thus hurting model performance based on our evaluation metrics. 

\subsubsection{ One-class support vector machines} Support vector machines (SVM)~\cite{cortes1995support} are supervised learning models for classification tasks. The objective of the algorithm is to maximize the distance of the decision boundary between the classes using multiple datapoints as supports. With a trained SVM, new examples can be mapped into the same high dimensional space and predicted based on the side where the new datapoint falls onto. One-class support vector machines (OCSVM)~\cite{manevitz2001one,chen2001one} is thus a special case of SVM where the algorithm learns the decision boundary that envelops the only class of the data. From an anomaly detection perspective, OCSVM learns the corresponding support vectors from the normal data. During inference time, new data that are not mapped inside the enveloped boundaries is therefore considered an anomaly. In an OCSVM model, user-defined hyperparameters include the upper-bound of the error fraction $\nu$, and the scale factor of the radial basis function (RBF) kernel coefficient $\gamma$. In our experiments, $\nu$ is set to 0.001 and $\gamma$ is set to $1/n_f$, where $n_f$ is the number of input features.

In this approach, the anomaly measure (AM) is defined as the distance to the hyperplane (i.e., the separating decision boundary). As normal classes are assigned a positive score, the distance to the hyperplane has a positive value. Anomaly points which lies outside of the decision boundary will have a negative distance. The scores are then normalized in the following manner to craft a decision threshold:
\[ \tilde{m}_{n,i} = \frac{m_{n,i}}{\text{max}(m_n)} \quad,\quad \tilde{m}_{e,i} = \frac{m_{e,i}}{\text{max}(m_n)}\]
where $\tilde{m}_{n,i},\tilde{m}_{e,i}$ denote the normalized anomaly measures for the normal data (with subscript $n$) and the evaluation data containing anomaly (with subscript $e$) respectively for the $i$-th datapoint, and $m_{n,i},m_{e,i}$ are the pre-normalized anomaly measures. Note that the anomaly measure of the evaluation set is normalized with respect to the maximum anomaly measure in the normal data used for training the model. Then, the decision threshold $t$ is defined to be three standard deviations away from the \textit{median} of the normal data: \[ t = \bar{m}_n - 3s_{m_n} \] where $\bar{m}$ is the median and $s_{m_n}$ is the standard deviation of the anomaly measure for normal data. Median is preferred over the mean as mean is greatly affected by the presence of false anomaly spikes even in the normal data. Anomaly measures that have been normalized for the anomalous data will present themselves as huge spikes way below the threshold $t$. Hence, these points are marked as attacks.

\subsubsection{ Isolation forests} Isolation forest~\cite{liu2008isolation} is an unsupervised outlier-anomaly detection framework that operates in a same manner as random forests~\cite{liaw2002classification}. In random forests for classification, the target class label is reached by appropriately splitting the values of the input features for making decisions similar to a decision tree. Multiple trees are learned and their predictions are averaged to form a strong random forest classifier. Instead of performing classification, isolation forests isolates the observation by randomly selecting a split value for randomly selected input features. The majority of the normal data will require lots of splits in the feature space to effectively isolate the points, but an anomalous data can be quickly isolated with very few feature splits since they are very dissimilar to normal data. Hence, the anomaly measure is defined as the average path length from the root node to the terminal node in isolating a particular point. In contrast to OCSVM, the anomaly measure is higher for normal data points, but very low for anomaly points. The anomaly measure is normalized and thresholded in the same manner as of the normalization scheme employed in OCSVM outlined above. Important hyperparameters include the number of trees of $n_{trees}=200$ and a contamination fraction of $f=10^{-8}$. Specifying the contamination fraction functions as slack since occasional spikes in anomaly measure are present even in normal data.

\subsubsection{ Synthetic corruption and random forests} This approach is similar to artificial anomaly association presented in~\cite{liu2016root} for root-cause anomaly detection and formulates the problem as a supervised classification task using random forests as the base classifier. In this formulation, normal data $X_{original}$ is given a label of `0' (i.e., a boolean equivalent indicating \textit{not attacked}). A new set of data $X_{corrupted}$ is derived from the normal data by randomly corrupting the normal data with a random noise matrix of $m$ rows and $n$ columns:
\[ X_{corrupted}^{m\times n} =  N^{m\times n}\circ X_{normal}^{m\times n} \quad,\quad N_{i,j}\sim\text{Uniform}(0,1) \]
where each row entry of the corrupted data $X_{corrupted}$ is given the label of `1' indicating corrupted data (i.e., input feature interdependencies are now broken). The operator $\circ$ indicates element-wise multiplication. Training proceeds as a binary classification problem in a supervised manner. During inference, the trained model is evaluated on the evaluation data where each individual test data points will be predicted as either `0' (normal) or `1' (attacked/anomalous). It is expected that this approach would work well for {\em disconnect} attacks.

\subsubsection{ Principal component analysis (PCA) with convex hulls} PCA~\cite{jolliffe2002principal} is a statistical procedure that converts a set of observations of correlated variables $X_{unprojected}$ into a set of linearly uncorrelated variables $X_{projected}$ named as principal components using an orthogonal transformation. Upon performing this transformation, the first resultant principal component has the largest possible variance that accounts for the maximum amount of variability in the data. One can then project features from a higher-dimensional space onto a lower-dimensional space before applying anomaly detection algorithms.

In this approach, a convex hull is constructed over the projected normal points $X_{projected}$ in the low-dimensional space. Mathematically, the convex hull is a set of points $X$ in the Euclidean space that is the smallest convex set that contains $X$. During inference, any points projected into coordinates outside the learned convex hull are deemed as anomalous. This method is easy to compute but not robust to extreme values. However, we observe great performance in the present problem and discussions will be presented in section~\ref{sec:results}.

\subsubsection{ Inverse-PCA technique} Similar to DAEs, inverse PCA relies on encoding information from a higher-dimensional space into lower-dimensional space within the principal components. A transformation matrix for forward PCA is computed from the normal data $X_{unprojected}$, and the data is reconstructed by performing inverse PCA transform using the same transformation matrix on the low dimensional data $X_{projected}$ to form $\hat{X}_{unprojected}$. The reconstruction error is computed by comparing the reconstructed data versus the original data. Ideally, the reconstruction error from a normal data will be low. However, when inverse PCA is applied on anomaly points, one would expect a huge spike in the reconstruction error $e$. Here, we used the mean-squared error (MSE) as a metric for the reconstruction error. Similar to discussed methods above, the decision threshold $t$ is set to three standard deviations away from the mean error observed in the normal data, i.e.: $t = \bar{e}_n + 3s_{e_n}$

\section{Results and discussion}
\label{sec:results}
\begin{table*}[htb!]
\caption{Algorithm performance for DAE, Neural Networks (NN), one-class support vector machines (OCSVM), isolation forest (IsoRF), corruption random forest (Corrupt RF), PCA with Convex Hull (PCA-CH) and inverse PCA (iPCA) for various attack modes. The overall attack covers the scenario where an attacker performs all four attacks in a single day. The best performer is selected by ranking the F1 score. These sets of metrics are presented in boldface.}
\vspace{5pt}
\label{tab:perf}
\begin{center}
\begin{tabular}{ccccccccc}
	&		&		DAE&  NN& OCSVM	&	IsoRF	&	CorruptRF	&	PCA-CH	&	iPCA	\\\hline
Overall	&	Precision	&	11.89&   43.60&	95.90	&	22.32	&	60.80	&	\textbf{70.02}	&	85.43	\\
	&	Recall / TPR	&		47.21 & 54.52 &57.12	&	54.62	&	70.18	&	\textbf{83.64}	&	34.43	\\
	&	F1	&		18.48& 47.65 & 71.60	&	31.69	&	65.16	&	\textbf{76.22}	&	49.08	\\
           &	Accuracy &	82.67& 95.14& 96.33& 80.95& 93.93&\textbf{95.78}& 94.22\\
	&	ROC-AUC	&	NA& NA&	78.45	&	68.94	&	83.10	&	\textbf{90.24}	&	66.96	\\\hline
Disconnect	&	Precision	&	13.58& 38.51&	95.63	&	20.35	&	\textbf{74.81}	&	73.31	&	85.71	\\
	&	Recall / TPR	&	73.21& 63.62&	51.98	&	49.08	&	\textbf{100.00}	&	100.00	&	36.41	\\
	&	F1	&		22.91& 47.98&67.35	&	28.77	&	\textbf{85.59}	&	84.60	&	51.11	\\
	&	Accuracy &	78.10& 93.87& 95.92& 80.34& \textbf{97.22} & 97.05& 94.36\\
	&	ROC-AUC	&	 NA&NA&	75.88	&	66.08	&	\textbf{98.49}	&	98.40	&	67.94	\\\hline
Reverse Power flow	&	Precision	&	9.56& 25.21 &	93.59	&	18.05	&	\textbf{61.39}	&	63.94	&	4.76	\\
	&	Recall / TPR	&		36.83& 19.64& 38.52	&	40.63	&	\textbf{100.00}	&	59.89	&	0.26	\\
	&	F1	&		15.18& 22.08& 54.58	&	25.00	&	\textbf{76.07}	&	61.85	&	0.50	\\
	&	Accuracy &	81.71&  93.84&  94.81&  80.27& \textbf{95.00}&  94.02&  91.50\\
	&	ROC-AUC	&		NA& NA& 69.15	&	62.20	&	\textbf{97.29}	&	78.46	&	49.90	\\\hline
Power Curtailment	&	Precision	&	10.73& 39.18&	94.12	&	13.02	&	55.38	&	\textbf{67.22}	&	14.81	\\
	&	Recall / TPR	&	18.30& 36.38& 	37.99	&	28.76	&	37.31	&	\textbf{74.67}	&	1.06	\\
	&	F1	&		13.53& 37.73& 54.14	&	17.93	&	44.58	&	\textbf{70.75}	&	1.97	\\
	&	Accuracy &	89.60& 94.66& 94.79& 78.69& 92.36& \textbf{95.00}& 91.50\\
	&	ROC-AUC	&	NA& NA&	68.89	&	55.92	&	67.30	&	\textbf{85.73}	&	50.26	\\\hline
VAR	&	Precision	&	13.68& 71.47&	\textbf{97.68}	&	34.27	&	44.02	&	73.17	&	94.28	\\
	&	Recall / TPR	&	60.49& 98.44& 	\textbf{100.00}	&	100.00	&	43.55	&	100.00	&	100.00	\\
	&	F1	&		22.31& 82.82& \textbf{98.83}	&	51.04	&	43.78	&	84.50	&	97.06	\\
	&	Accuracy &	81.28& 98.18& \textbf{99.81}& 84.48& 91.12& 97.03& 99.51\\
	&	ROC-AUC	&	NA& NA&	\textbf{99.90}	&	91.56	&	69.39	&	98.39	&	99.73	\\

\end{tabular} 
\end{center}
\end{table*}

This section discusses the performance of our fused anomaly detection framework, machine learning algorithms and also the performance of $m_3$ model to detect large scale attacks. We provide various insights on the results obtained. Our simulations assume that PVs are deployed far away from the substation and  $100\%$  of houses/PVs in a randomly chosen residential area are attacked. Each house has a multi-modal anomaly detection technique that monitors PV, house load and node measurements to detect attacks.
\subsection{Performance metrics}
We evaluate the performance of the models based on precision, recall and F1 score. In the context of classification, precision (Pr) and recall (Re) encapsulates the true positives (TP), true negatives (TN), false positive (FP), and false negatives (FN) of the model. The terms \textit{true} and \textit{false} denote whether the model's prediction corresponds to an external judgement, whereas the terms \textit{positive} and \textit{negative} refer to the prediction of the classifier. 
Then, precision is defined as:~\mbox{$Pr=\frac{TP}{TP+FP}$} and the recall, equivalent to true positive rate (TPR) is defined as:~\mbox{$Re=\frac{TP}{TP+FN}$}.

The F1-score metric is defined as: \mbox{$F1 = \frac{2{Pr}{Re}}{{{Pr} + {Re}}}$}, which is a single metric computed as the harmonic mean between precision and recall. 
The receiving operator characteristics (ROC) curve plots the true positive rate against the false positive rate. Ideally, a perfect classification model will have an area under the ROC curve (alternatively named as AUROC, or ROC) as 1, whereas a model that is guessing randomly will have an ROC of around 0.5. The larger the value, the more favorable is the model's performance.
\subsection{Performance of machine learning algorithms}
\begin{figure*}[htbp]
\centering
% disconnect attack
\begin{subfigure}{0.4\textwidth}
  \centering
  \includegraphics[width=7cm]{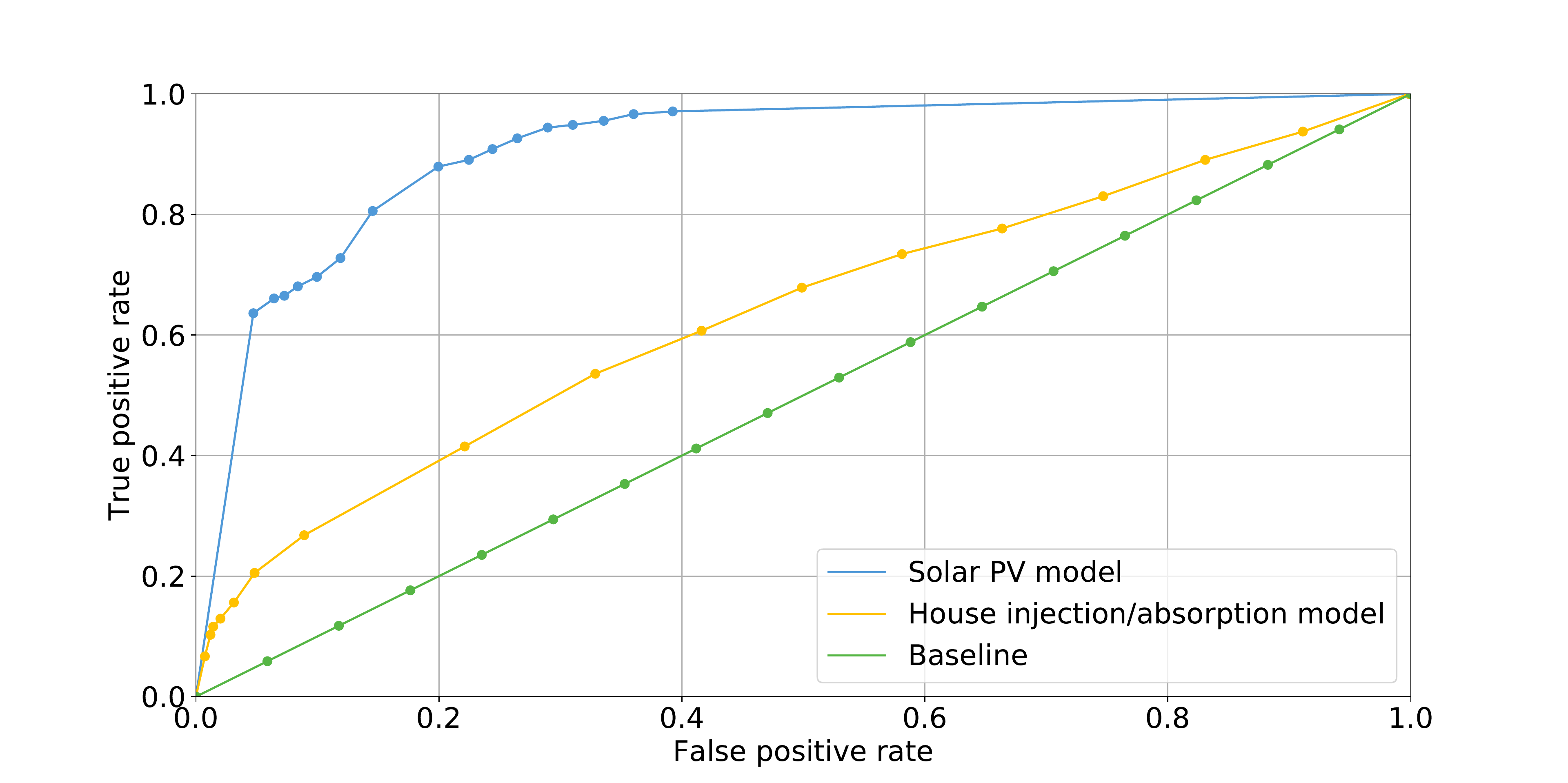}
  \caption{PV disconnect attack}
  \label{figure_disconnect}
\end{subfigure} \hspace{0in}
% curtailment attack
\begin{subfigure}{0.4\textwidth}
  \centering
  \includegraphics[width=7cm]{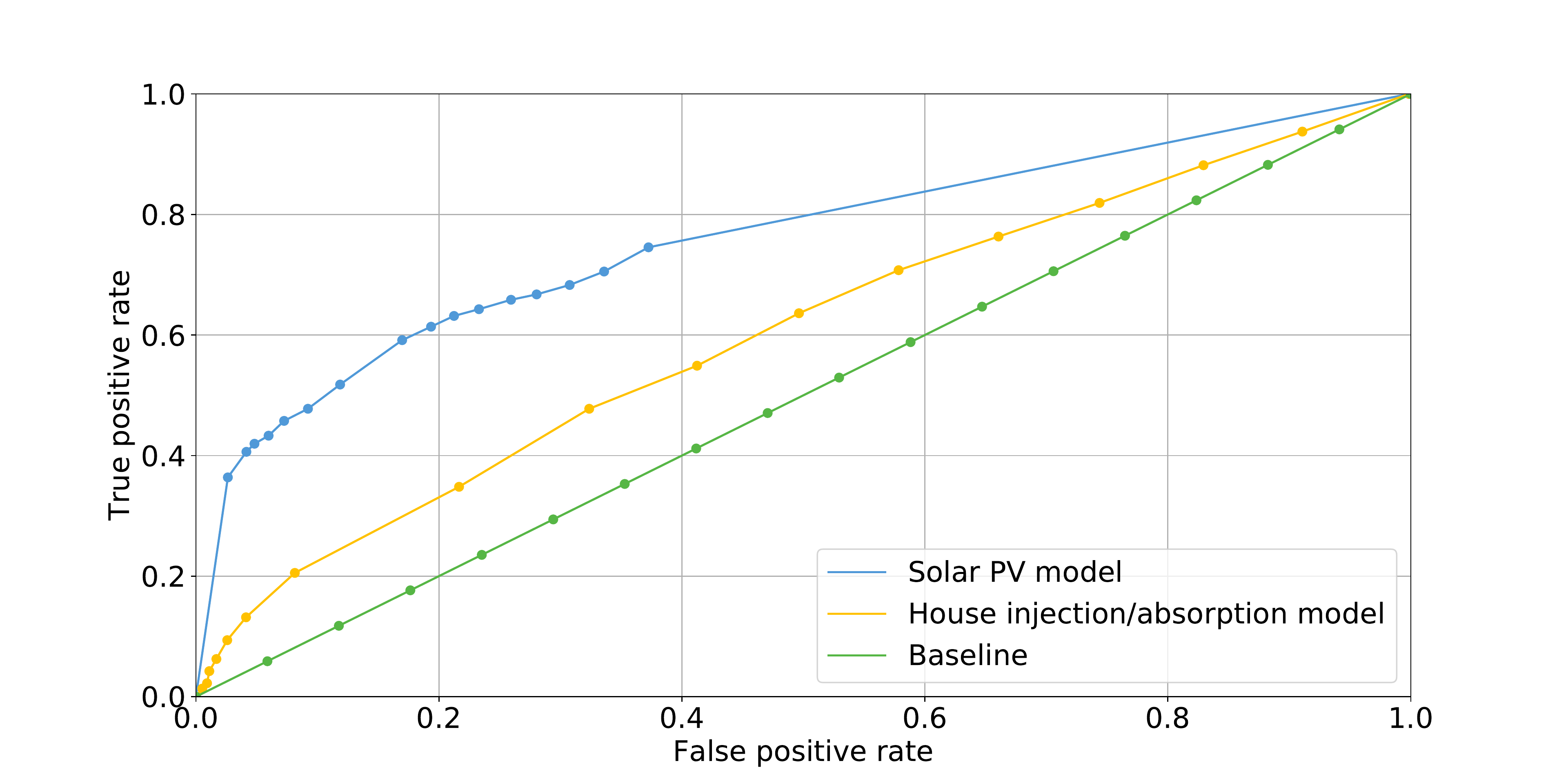}
  \caption{Power curtailment attack}
  \label{figure_curtailment}
\end{subfigure}
\\ % var attack
\begin{subfigure}{0.4\textwidth}
  \centering
  \includegraphics[width=7cm]{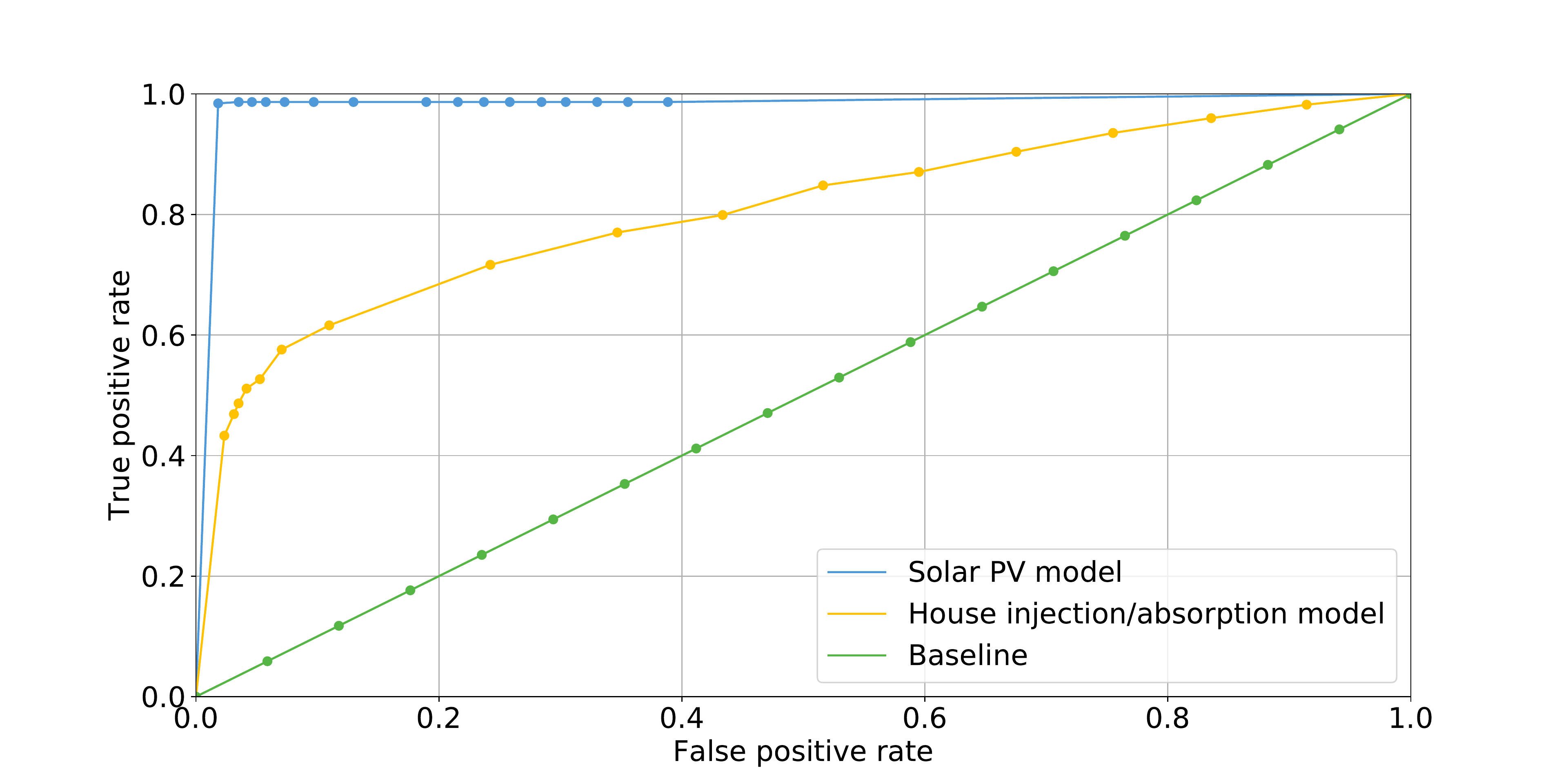}
  \caption{Volt-var attack}
  \label{figure_var}
\end{subfigure} \hspace{0in}
% load disconnect attack
\begin{subfigure}{0.4\textwidth}
  \centering
  \includegraphics[width=7cm]{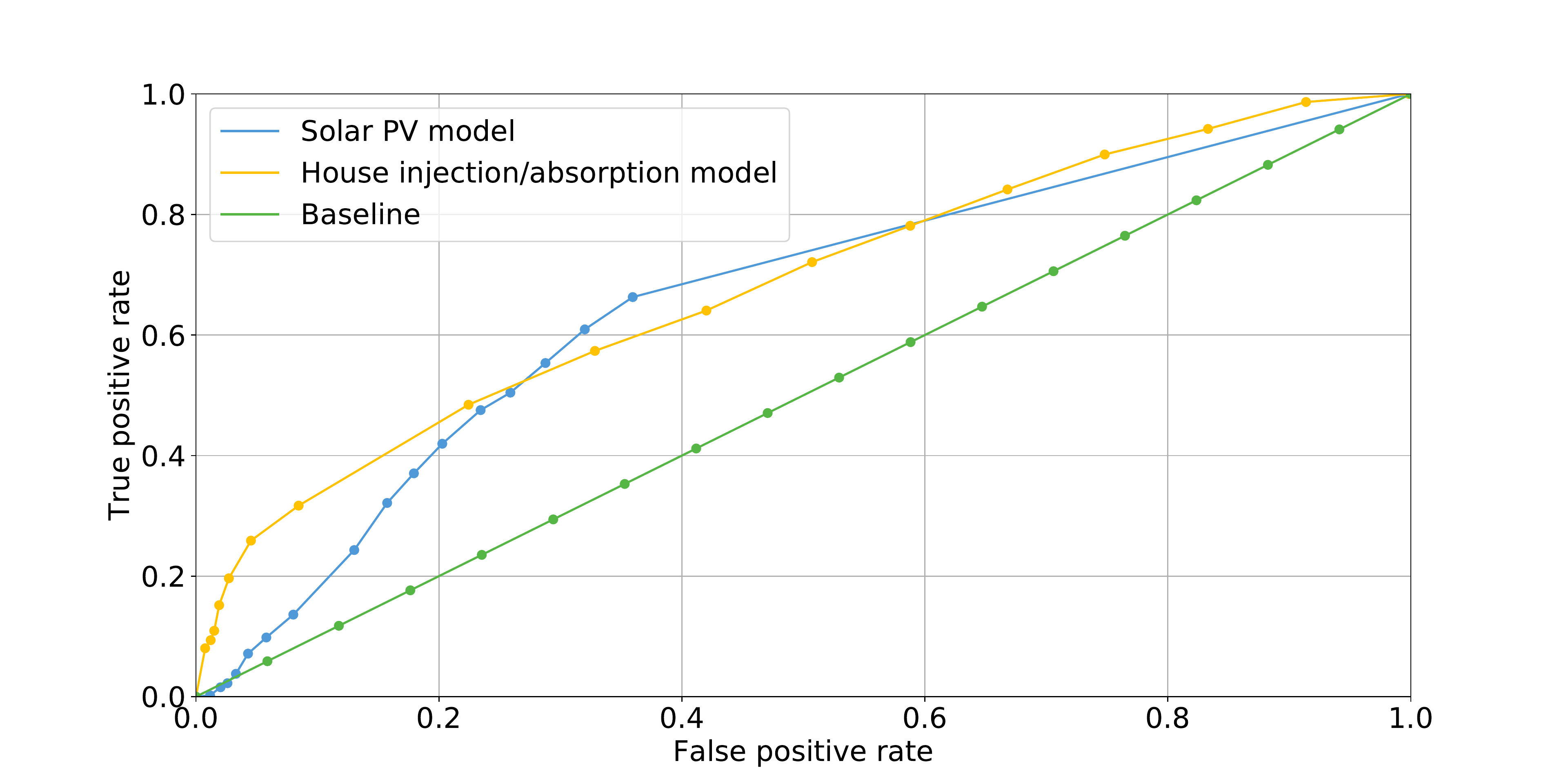}
  \caption{Reverse power flow attack}
  \label{figure_power}
\end{subfigure}
\caption{ROC curves based on Neural Networks for PV and house load model under four attack modes}
\label{figure_detecte_solar}
\end{figure*}

\begin{figure*}[htbp]
\centering
% disconnect attack
\begin{subfigure}{0.4\textwidth}
  \centering
  \includegraphics[width=7cm]{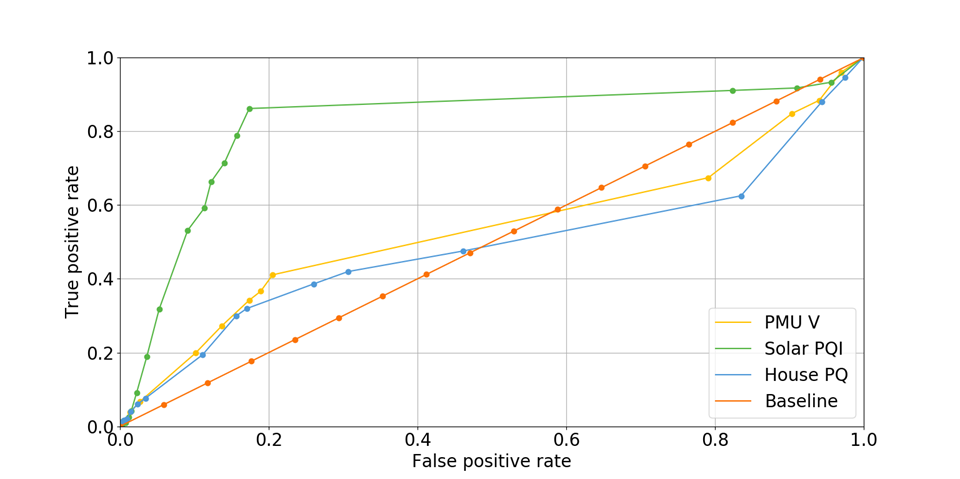}
  \caption{PV disconnect attack}
  \label{figure_disconnect}
\end{subfigure} \hspace{0in}
% curtailment attack
\begin{subfigure}{0.4\textwidth}
  \centering
  \includegraphics[width=7cm]{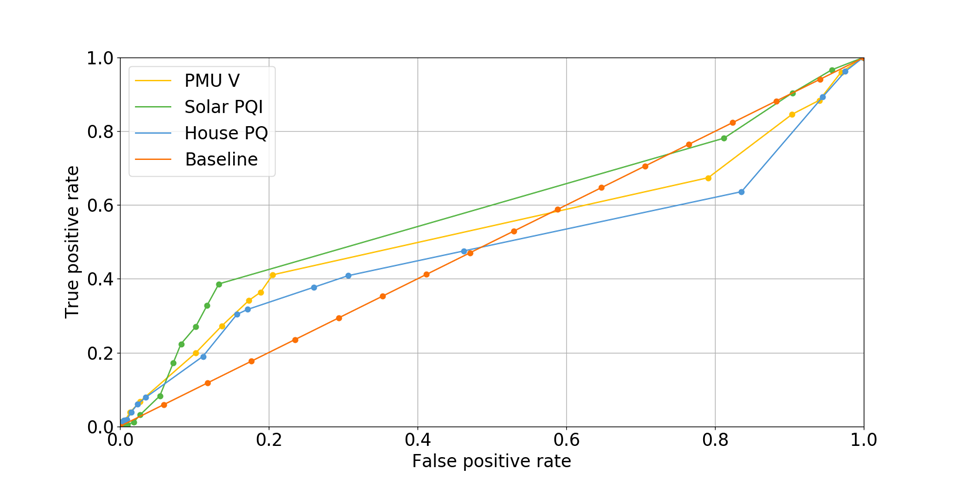}
  \caption{Power curtailment attack}
  \label{figure_curtailment}
\end{subfigure}
\\ % var attack
\begin{subfigure}{0.4\textwidth}
  \centering
  \includegraphics[width=7cm]{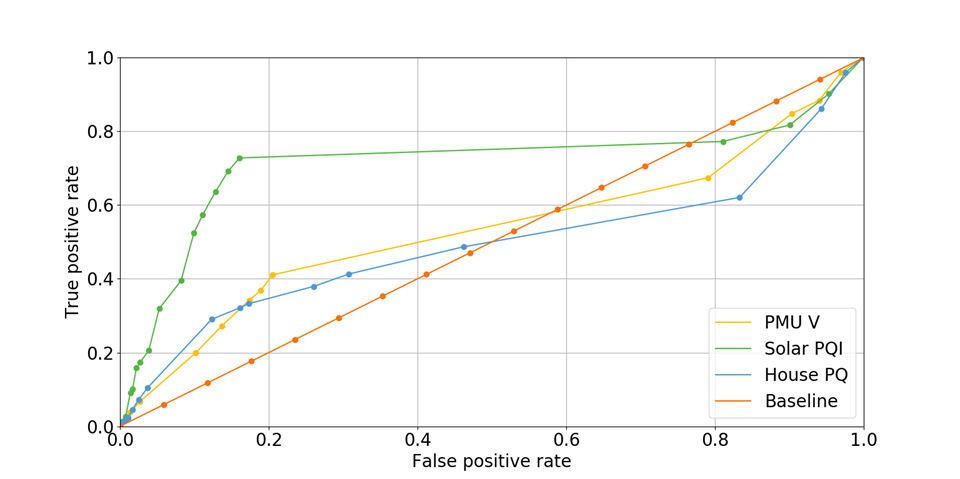}
  \caption{Volt-var attack}
  \label{figure_var}
\end{subfigure} \hspace{0in}
% load disconnect attack
\begin{subfigure}{0.4\textwidth}
  \centering
  \includegraphics[width=7cm]{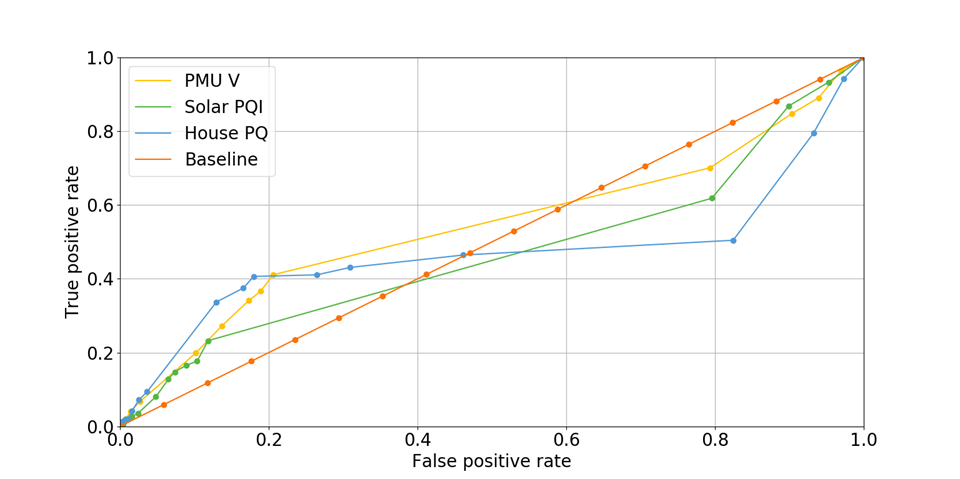}
  \caption{Reverse power flow attack}
  \label{figure_power}
\end{subfigure}
\caption{ROC curves based on DAEs for PV, house load and node model under four attack modes}
\label{figure_detecte_solar_ae}
\end{figure*}

 Figures \ref{figure_detecte_solar} and \ref{figure_detecte_solar_ae} depict the ROC curves for neural network and DAE algorithms, respectively, with regard to the four attacks. Following observations are made from the graphs: (i) Neural networks outperform DAE in the detection of all attacks; (b) For a false positive rate (FPR) of 0.2, using neural network, {\em volt-var} and {\em disconnect} attacks are detected with true positive rates (TPR) of 0.98 and 0.86 respectively; (c) As opposed to {\em disconnect} attacks, {\em power curtailment} attacks are detected with a TPR of $<0.6$ for a FPR of 0.2; (d) PV model outperforms house and node models for {\em disconnect}, {\em power curtailment} and {\em volt-var} attack as these attacks directly impact the PV active and reactive power and can be easily detected; (e) House model slightly performs better compared to PV model in the event of a {\em reverse power flow} attack, however, generally, the performance of house and node models are worse compared to PV model which mainly stems from the fluctuating house load demands and power generation.  
\vspace{5pt}\\

\begin{figure*}[h]
    \centering
    \includegraphics[width=0.8\textwidth,trim={0 100 0 0}]{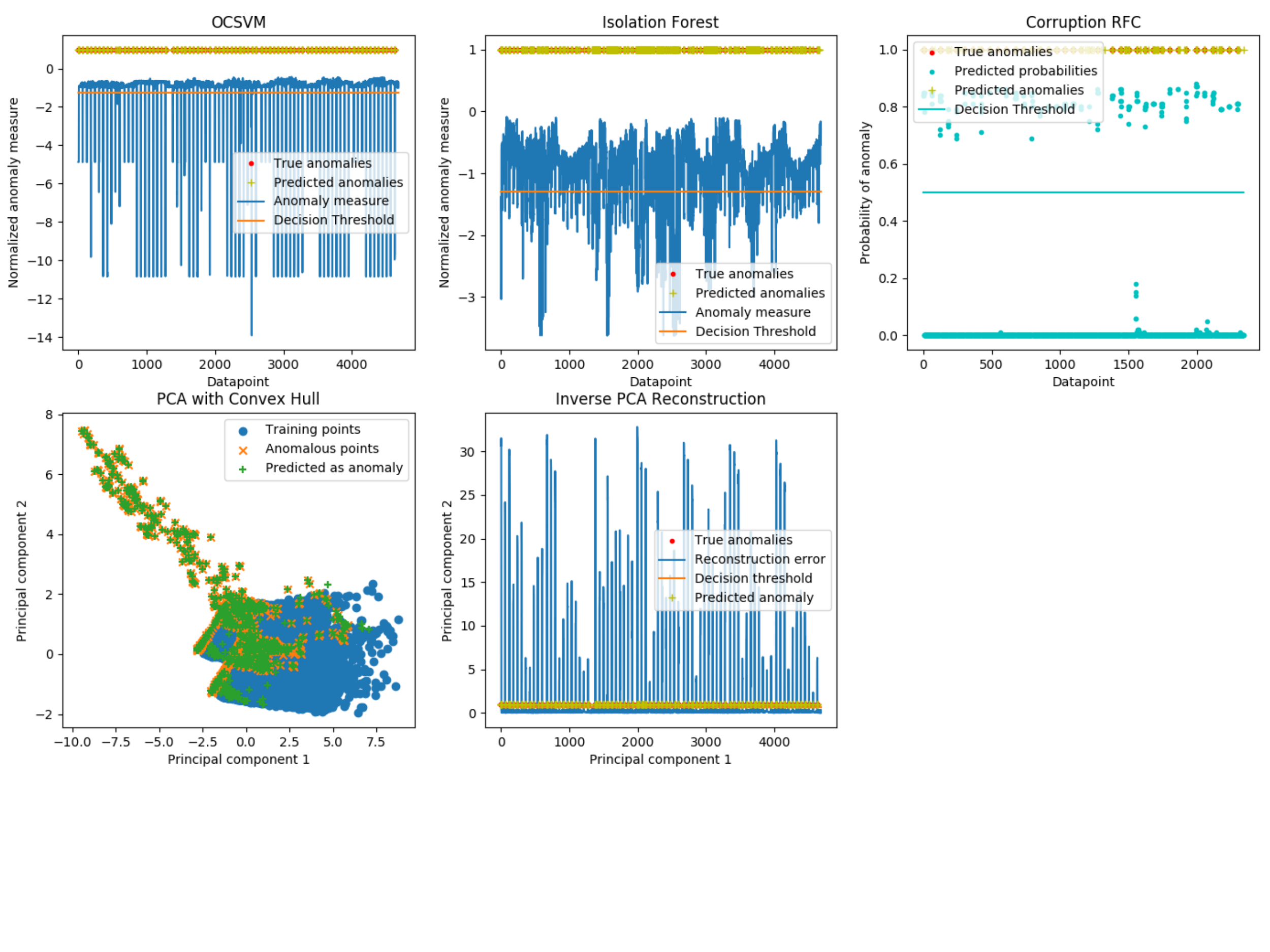}
    \caption{Illustration on how decisions are arrived using different anomaly detection frameworks. From left-to-right, top-to-bottom: (i) One class SVMs, (ii) Isolation Forests, (iii) Corruption RFC, (iv) PCA with convex hull, and (v) inverse PCA reconstruction. (i) and (ii): Normalized anomaly measures appear as spikes that greatly falls below the decision threshold. (iii): High confidence of anomaly class for anomalous points using Random Forests. (iv): Visualization of PCA components in 2 dimensions. In reality, the feature dimension is actually reduced to 5 dimensions to construct a convex hull. Points marked green are the predictions; if it coincides the orange points, it is a correct anomaly identification. (v): Reconstruction errors from inverse PCA. Similarly, anomalies manifest themselves as spikes in reconstruction errors.}
    \label{figure:anomalymethods}
\end{figure*}

\begin{figure*}[h]
    \centering
    \includegraphics[width=0.7\textwidth,trim={48 32 48 0}]{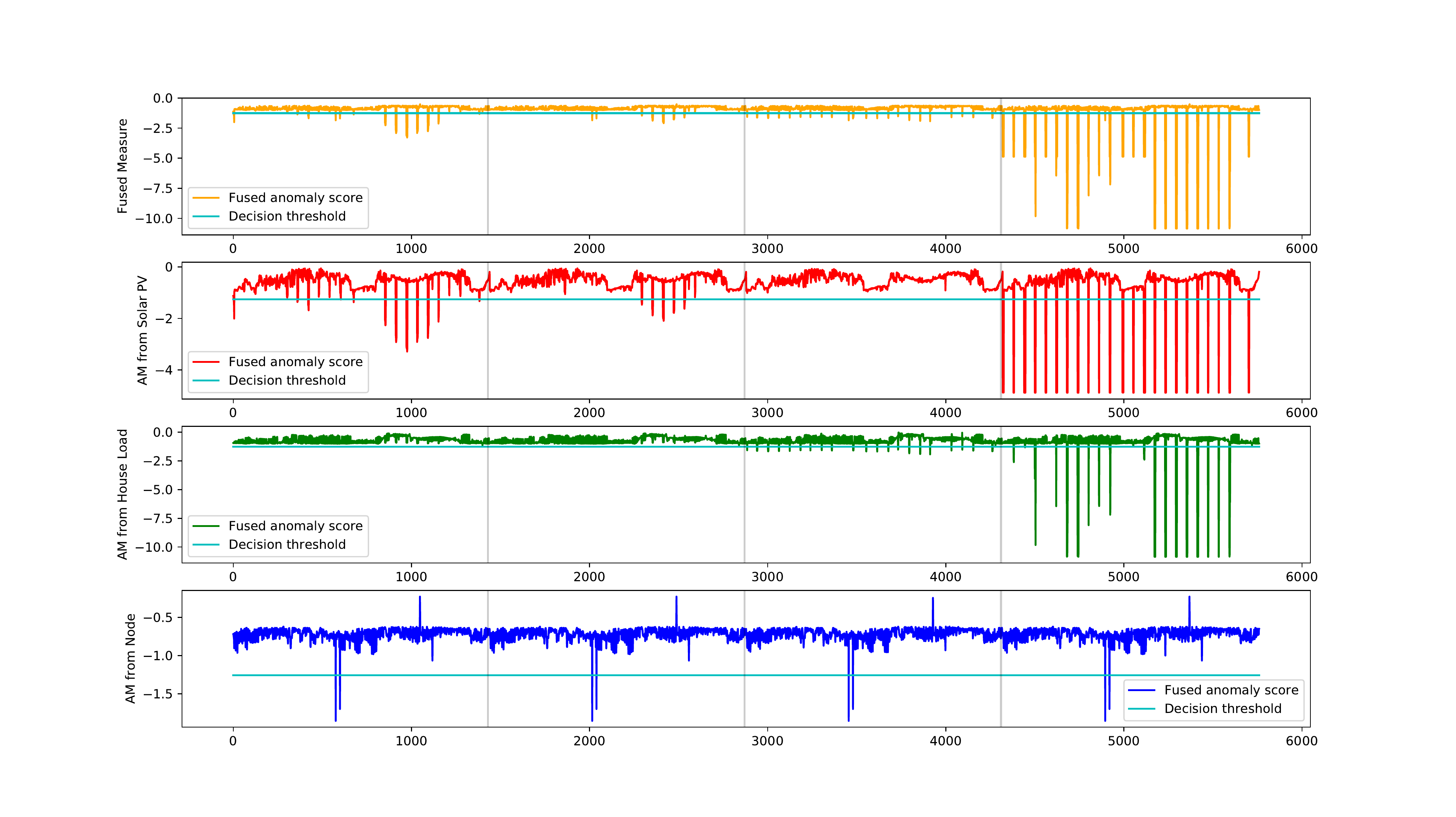}
    \caption{Fused anomaly measure under the \textit{MostAnomalous} fusion scheme for a subset of the time series data using OCSVM as the anomaly detector. Individual performance of the models are also illustrated }
    \label{figure:fusedscores}
\end{figure*}

Table \ref{tab:perf} depicts the performance of (1)-(7) classifiers (DAE, neural networks, one-class support vector machines, isolation forest, corruption random forest, PCA with Convex Hull and inverse PCA) for various attacks. The results presented for each algorithm is based on the {\em MostAnomalous} fused measure. Figure~\ref{figure:anomalymethods} shows on how decisions are generated for various anomaly detection method frameworks. 
From the experiments, we observe that each algorithm excels in identifying a particular attack type. Based on the overall performance, PCA-CH is the most robust to the four attack types. As most attacks directly result in a huge disturbance to the extent of distorting feature correlations, the attacked instance lies in a feature space --- outside the convex hull of normal points --- that is not anticipated by the trained model. PCA-CH appears to perform this task well, followed by OCSVM, the second-best performer accomplishes a similar task by constructs a separating hyperplane to distinguish normal data from unexpected data. One can further examine individual attack types to study the strengths and weaknesses of each model. For {\em disconnect} and {\em reverse power flow} attacks, random forest classification with synthetic anomaly injection is the best performer with very competitive performance compared to PCA-CH, the overall best model. This result is not surprising because the training approach of noise injection inherently captures the ``statistics of sudden disconnects''. For {\em power curtailment} attacks (i.e. denoted by power), PCA-CH remains unchallenged as it can still effectively detect broken correlations between measured variables. Interestingly, attacks on reactive power can be most effectively determined by the simpler OCSVM model at the house level which considered only two features --- active and reactive power of house --- which exhibits clear disparity in correlation whenever attacks occur. It is worth noting that this simple correlation break can also be captured quite well by other algorithms, such as NN, PCA-CH, and iPCA.
\vspace{5pt}\\
Using precision and recall-based metrics for evaluation has its limitations. At every time instant, an anomaly decision is made using the set of measurements obtained at that time instant and hence favors point-wise classification approaches such as OCSVM, IsoRF, CorruptRF, PCA-CH, and iPCA. On the other hand, our time series formulation with DAE and neural networks suffer as they were only able to detect the instant where attack occurs. For example, the first instant of an attack of arbitrary temporal length can be detected by the algorithm but the decision on the second time instant is affected by performing regression using anomalous features. An event-based metric is desired for more accurate analysis; unfortunately, there is no standard definition of such a metric in literature. Hence, only precision and recall-based metrics are employed for reporting consistency.
\begin{figure}[htbp!]
    \centering
    \includegraphics[width=1\columnwidth]{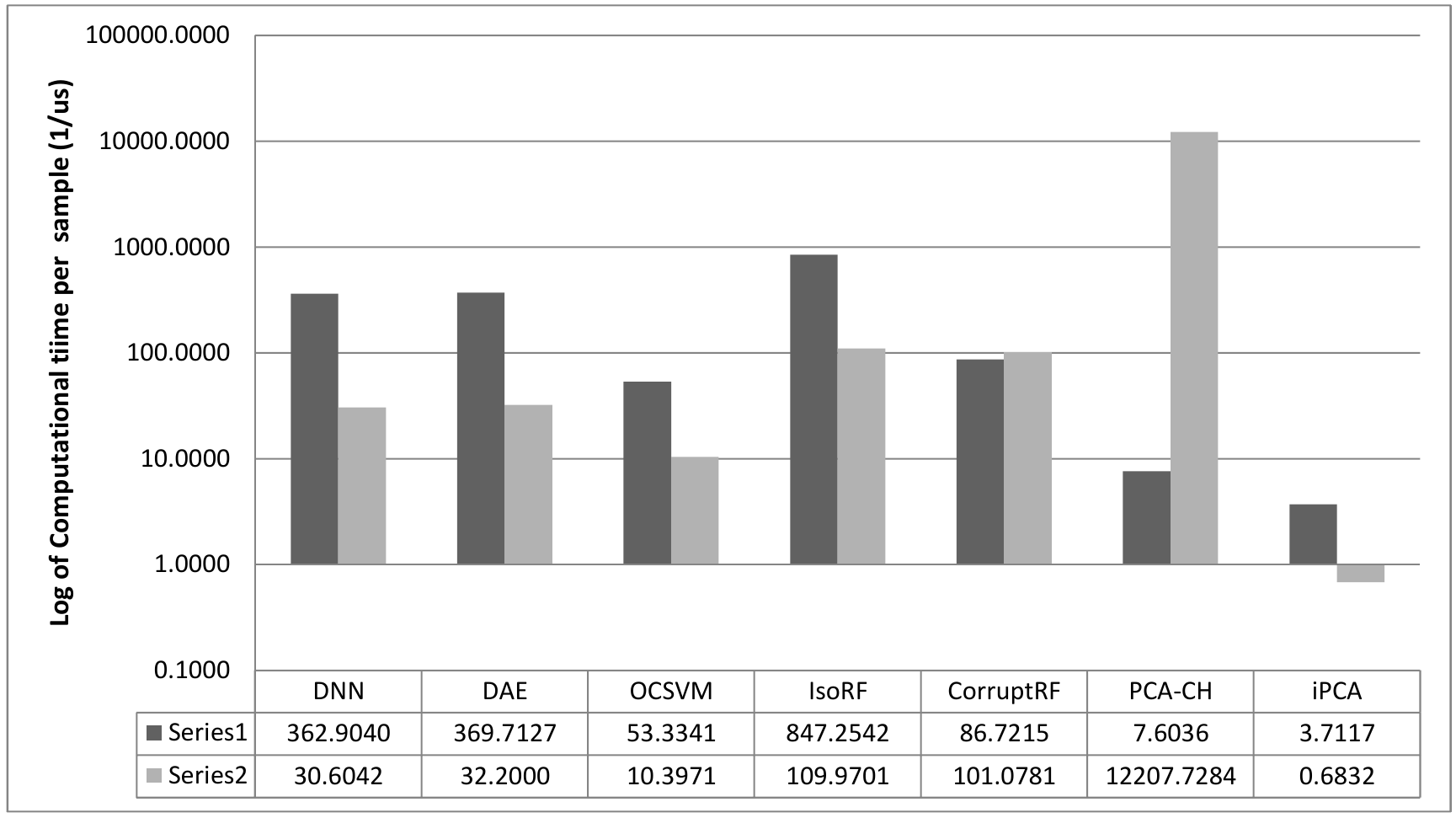}
    \caption{Computational time of training and testing machine learning algorithms per sample in  microseconds. The series $1$ and $2$, respectively,  represent the training and testing time in samples per microseconds.}
    \label{figure:ctime}
\end{figure}
The computational time of the algorithms is presented in Figure~\ref{figure:ctime}. Among all algorithms, iPCA requires the least time to train and test on one sample, followed by OCSVMs and CorruptRF classification. As a classification algorithm, these approaches are expected to have a faster computation time compared to regression approaches using NN and DAE. The deep learning methods have a slightly higher execution time with most of the time spent on training. On the other hand, Isolation Forest constructs numerous decision pathways to determine the average path lengths as a measure of anomaly and it has to be done separately for each data points. Hence, we observe a significant increase in computation time. One interesting result is that while the performance of PCA-CH is consistently robust over all four attack types, it is also the algorithm that requires the longest time to run. It is straightforward to determine the convex hull of the training points but determining whether a single test instance lies within the higher dimension hull requires performing a more complex procedure called the Delaunay triangulation. Based on the criteria ``that balances between both execution time and anomaly detection capabilities'', OCSVM and CorruptRF are two of the best algorithms.
\vspace{5pt} \\
\noindent{\em Performance of fused anomaly measure}: The performance of the fused anomaly measure under the \textit{MostAnomalous} fusion scheme for a subset of the time series data using OCSVM as the anomaly detector is shown in Figure~\ref{figure:fusedscores}. The first plot shows the fused anomaly measures for four attack types (i.e., \textit{disconnect, power curtailment, reverse power flow, volt-var}) partitioned by gray vertical lines. The second plot shows the anomaly measure obtained from the PV model, the third from load model, and the fourth from the node model. Our PV model is able to reliably detect \textit{disconnect, power curtailment} and \textit{volt-var} attacks based on the automatically-determined decision threshold whereas the detection capability for \textit{reverse power flow} attacks is complemented by the load model.
\subsection{Performance of $m_3$ model for large scale attack detection}
We run simulations for different attack penetration levels (i.e., $25\% - 100\%$ of houses/PVs in a residential area are attacked) for \textit{disconnect, power curtailment, reverse power flow} and \textit{volt-var} attacks. Table \ref{figure_detect_pmu} presents the performance of node model ($m_3$) based on neural networks and DAE for detecting attacks. As the number of houses/PVs attacked increases, the detection performance of our algorithm increases. However, the reduced detection performance (40\%) is due to the following two reasons: (a) we obtain the voltage magnitude and phase angle from GridLAB-D measurements every 1 minute which in turn limits our ability to capture the voltage transients effectively; (b) heterogeneous nature of load supply and demands also cause the nominal voltage to fluctuate in normal cases.  We plan to improve the performance of our node model using two techniques: (i) acquire data at a faster sampling rate to capture the voltage transients; (ii) use physics based features that captures the correlation between the load generation/consumption and feeder measurements (e.g., when active power is injected into the grid during the {\em reverse power flow} attack, feeder voltage magnitude and phase angle rises) to improve the model performance. 
\begin{table}[h!]
\caption{Performance of node model based on neural networks (NN) and auto-encoders (DAE) for varying attack penetration levels (25\% to 100\%) and attack modes. The results show the true positive rate when the false positive is around 0.2}
\vspace{5pt}
\label{figure_detect_pmu}
\scriptsize
\begin{tabular}{cccccc}
	&		&	 Disconnect  &   Curtailment & Volt-var&  Reverse power\\\hline
100\%&	NN	&	0.4 & 0.38 & 0.4 & 0.45	\\
	&	DAE	&	0.38 & 0.35 & 0.36& 0.35	\\\hline
75\%	&	NN	&	0.38& 0.36& 0.38& 0.39 	\\
	&	DAE	&	0.38& 0.35& 0.36& 0.35	\\\hline
50\%	&	NN	&	0.34& 0.32& 0.37& 0.32	\\
	&	DAE	&	0.35& 0.34& 0.35& 0.34		\\\hline
25\%	&	NN	&	0.31& 0.30& 0.37& 0.30	\\
	&	DAE	&	0.34& 0.34& 0.35& 0.34		\\
\end{tabular} 
\end{table}
\section{Conclusion}
\label{sec:conclusion}
In this work, we design and develop an edge-based distributed detection framework that continuously monitors PV and its interaction with the grid in real time to predict the presence of an attack. This is realized by using data-driven techniques on various sources of information from PV (active power, reactive power, voltage-current characteristics), smart meters (amount of power drawn and exported back to the grid), and $\mu$PMUs (feeder voltage magnitude and phase angles) and then a {\em MostAnomalous} fusion based scheme that fuses multiple observations and flags anomaly when a deviation from normal behavior is observed. Various attacks targeting PV panels were designed and performed on GridLAB-D. Our simulation results show that each algorithm excels in detecting a particular attack type (e.g., OCSVM detects {\em volt-var} attack with ROC-AUC of 99.9\% and CorruptRF detects {\em reverse power flow} and {\em disconnect} attacks with ROC-AUC of 97.27\% and 98.49\% respectively) and PCA with Convex Hull demonstrated to be the most robust algorithm to all four attack types with a F1 score and ROC-AUC of 76.22\% and 90.24\% respectively. 
%We find that {\em reverse power flow} attacks can be identified only by analyzing the power injected into and absorbed from the %grid, which is gathered from the smart meters. Similarly, feeder based measurements prove to be an useful measure to detect %large scale attacks on DERs.
In future, we propose to augment the existing statistical features with physics based features to accurately capture the correlation between the device and the grid to improve the performance of the model.  
\section*{Acknowledgment}
This material is based upon work supported by the Department of Energy under Award Number(s) DE-OE0000826.
\section*{Disclaimer}
This paper was prepared as an account of work sponsored by an agency of the United States Government.  Neither the United States Government nor any agency thereof, nor any of their employees, makes any warranty, express or implied, or assumes any legal liability or responsibility for the accuracy, completeness, or usefulness of any information, apparatus, product, or process disclosed, or represents that its use would not infringe privately owned rights.  Reference herein to any specific commercial product, process, or service by trade name, trademark, manufacturer, or otherwise does not necessarily constitute or imply its endorsement, recommendation, or favoring by the United States Government or any agency thereof.  The views and opinions of authors expressed herein do not necessarily state or reflect those of the United States Government or any agency thereof.

\bibliographystyle{abbrv}
{
\bibliography{references}
} 

\end{document}